\documentclass[review]{elsarticle}
\usepackage[margin=3cm]{geometry}

\usepackage{lineno,hyperref}

\usepackage{caption}
\usepackage{subcaption}
\usepackage{acro}
\acsetup{first-style=short}
\usepackage{dblfloatfix}
\usepackage{amsmath}
\usepackage[table]{xcolor}
\usepackage{graphicx}
\graphicspath{{./CAS_IJMF}}
\usepackage{booktabs}
\usepackage{siunitx}
\usepackage[utf8]{inputenc}
\usepackage{upquote}
\usepackage{csquotes}
\usepackage{ragged2e}

\usepackage{newtxmath}

\journal{International Journal of Multiphase Flow}

\biboptions{authoryear}

\usepackage{pgfplots}

\newcommand{\Lnoz}{\mathsf{L_{noz}}}

\newcommand{\Dnoz}{\mathsf{D_{noz}}}
\newcommand{\Do}{\mathsf{D_o}}
\newcommand{\Uo}{\mathsf{U_o}}
\newcommand{\tauo}{\mathsf{\tau_o}}

\newcommand{\Tb}{{T_{\mathrm{b}}}}
\newcommand{\Tg}{{T_{\mathrm{g}}}}

\newcommand{\Red}{\mathsf{Re}_{\mathrm{{d}}}}
\newcommand{\BMd}{\mathsf{B_{M,\mathrm{{d}}}}}
\newcommand{\Shd}{\mathsf{Sh_{\mathrm{{d}}}}}
\newcommand{\Scg}{\mathsf{Sc_{\mathrm{{g}}}}}
\newcommand{\Prg}{\mathsf{Pr_{\mathrm{{g}}}}}
\newcommand{\Nud}{\mathsf{Nu_{\mathrm{{d}}}}}

\newcommand{\Ad}{{A_{\mathrm{d}}}}
\newcommand{\Ts}{{T_{\mathrm{s}}}}
\newcommand{\Thats}{\hat{T}_{\mathrm{s}}}

\newcommand{\mean}[1]{\langle #1 \rangle}
\newcommand{\bigmean}[1]{\left\langle #1 \right\rangle}

\newcommand{\wenta}{\dot{\omega}_{\mathrm{ent,\air}}}

\newcommand{\wdrag}{\dot{\omega}_{\mathrm{drag}}}
\newcommand{\wvapextotal}{\dot{\omega}_{\mathrm{vap,ext}}}
\newcommand{\wvapswl}{\dot{\omega}_{\mathrm{vap,exp}}}
\newcommand{\wvapint}{\dot{\omega}_{\mathrm{vap,ht1}}}
\newcommand{\wvapext}{\dot{\omega}_{\mathrm{vap,ht2}}}
\newcommand{\wbre}{\dot{\omega}_{\mathrm{bre}}}

\newcommand{\wheat}{\dot{\omega}_{\mathrm{heat}}}
\newcommand{\gas}{\mathrm{g}}
\newcommand{\air}{\mathrm{a}}
\newcommand{\liq}{\mathrm{l}}
\newcommand{\vap}{\mathrm{v}}

\newcommand{\Dg}{\mathrm{D_g}}
\newcommand{\Dl}{\mathrm{D_l}}
\newcommand{\rhohat}{\bar{\rho}}
\newcommand{\rhol}{\rho_{\liq}}
\newcommand{\rhog}{\rho_{\gas}}
\newcommand{\rhoa}{\rho_{\air}}
\newcommand{\rhov}{\rho_{\vap}}
\newcommand{\vl}{v_{l}}
\newcommand{\Yhatl}{\hat{Y}_{\liq}}
\newcommand{\Yhatg}{\hat{Y}_{\gas}}
\newcommand{\Yhata}{\hat{Y}_{\air}}
\newcommand{\Yhatv}{\hat{Y}_{\vap}}

\newcommand{\Td}{T_\mathrm{d}}
\newcommand{\Rp}{R_\mathrm{p}}
\newcommand{\Pv}{P_{\vap}}
\newcommand{\Pg}{P_{\gas}}

\newcommand{\uhatl}{\hat{u}_{\liq}}
\newcommand{\ul}{{u}_{\liq}}
\newcommand{\uhatg}{\hat{u}_{\gas}}
\newcommand{\ug}{{u}_{\gas}}

\newcommand{\mug}{\mu_{\gas}}

\newcommand{\mul}{\mu_{\liq}}

\newcommand{\dd}{{d_\mathrm{d}}}
\newcommand{\ddhat}{\hat{d_\mathrm{d}}}
\newcommand{\ddhatsq}{\hat{d}^2_{\mathrm{d}}}

\newcommand{\ddsq}{{d}^2_{\mathrm{d}}}

\newcommand{\dstthm}{d_{\mathrm{st,thm}}}
\newcommand{\dstaero}{d_{\mathrm{st,aero}}}

\newcommand{\taubthm}{\tau_{\mathrm{b,thm}}}
\newcommand{\taubaero}{\tau_{\mathrm{b,aero}}}

\newcommand{\dstKH}{d_{\mathrm{st,KH}}}
\newcommand{\taubKH}{\tau_{\mathrm{b,KH}}}
\newcommand{\dstRT}{d_{\mathrm{st,RT}}}
\newcommand{\taubRT}{\tau_{\mathrm{b,RT}}}

\newcommand{\Nbub}{N_{\mathrm{bub}}}
\newcommand{\Vdrop}{V_{\mathrm{l}}}
\newcommand{\Rdot}{\dot{R}_\mathrm{b}}
\newcommand{\Thatd}{\hat{T}_{\mathrm{d}}}
\newcommand{\Tref}{T_{\mathrm{ref}}}

\newcommand{\Tamb}{T_{\mathrm{amb}}}
\newcommand{\Thatg}{\hat{T}_{\gas}}
\newcommand{\Thata}{\hat{T}_{\air}}
\newcommand{\Thatv}{\hat{T}_{\vap}}
\newcommand{\Cpl}{C_{\mathrm{p,\liq}}}
\newcommand{\Cpg}{C_{\mathrm{p,\gas}}}
\newcommand{\Cpa}{C_{\mathrm{p,\air}}}
\newcommand{\Cpv}{C_{\mathrm{p,\vap}}}
\newcommand{\Kbre}{K_{\mathrm{bre}}}

\newcommand{\Kbreaerod}{K_{\mathrm{bre,aero}}}
\newcommand{\Kbrethm}{K_{\mathrm{bre,thm}}}

\newcommand{\Kvapsup}{K_{\mathrm{vap,ext}}}
\newcommand{\Kvapsub}{K_{\mathrm{vap,ext}}}
\newcommand{\Kvapint}{K_{\mathrm{vap,ht1}}}
\newcommand{\Kvapext}{K_{\mathrm{vap,ht2}}}
\newcommand{\Kvapswl}{K_{\mathrm{vap,exp}}}
\newcommand{\Gvg}{\mathit{\Gamma}_{\mathrm{v,g}}}
\newcommand{\Kheatsup}{K_{\mathrm{heat}}}
\newcommand{\Kheatsub}{K_{\mathrm{heat}}}

\newcommand{\Cl}{C_{\liq}}
\newcommand{\lambdag}{\lambda_{\gas}}

\newcommand{\Cdrag}{C_{\mathrm{drag}}}

\newcommand{\rhostar}{\rho^{*}}
\newcommand{\Ystarl}{{Y}^{*}_{\liq}}
\newcommand{\Ystara}{{Y}^{*}_{\air}}
\newcommand{\Ystarv}{{Y}^{*}_{\vap}}

\newcommand{\ustarl}{{u}^{*}_{\liq}}
\newcommand{\ustarg}{{u}^{*}_{\gas}}

\newcommand{\ddstar}{{d^*_\mathrm{d}}}

\newcommand{\Tstard}{{T}^*_{\mathrm{d}}}
\newcommand{\Tstarg}{{T}^*_{\gas}}
\newcommand{\zstar}{{z}^*}
\newcommand{\tstar}{{t}^*}
\newcommand{\bstar}{{b}^*}

\newcommand{\DeltaT}{\Delta T}
\newcommand{\DeltaTstar}{\Delta T^*}

\newcommand{\isooct}{\textit{iso}-octane}
\newcommand{\isobut}{\textit{iso}-butanol}
\newcommand{\trid}{\textit{n}-tridecane}
\newcommand{\isohex}{\textit{iso}-hexane}
\newcommand{\pen}{\textit{n}-pentane}
\newcommand{\hexane}{\textit{n}-hexane}
\newcommand{\but}{\textit{n}-butanol}

\newcommand{\Pinj}{P_{\mathrm{inj}}}
\newcommand{\Tinj}{T_{\mathrm{inj}}}

\newcommand{\COtwo}{\mathrm{CO_2}}
\newcommand{\NOx}{\mathrm{NO_x}}

\newcommand{\epcrit}{\epsilon_{\mathrm{crit}}}

\newcommand{\myeq}[1]{Eq.~\eqref{#1}}
\newcommand{\myeqs}[1]{Eqs.~\eqref{#1}}
\newcommand{\myfig}[1]{Fig.~\ref{#1}}
\newcommand{\mytab}[1]{Table~\ref{#1}}
\newcommand{\mysec}[1]{Section~\ref{#1}}

\definecolor{SAEyellow}{rgb}{1,0.698039215686274,0.00392156862745098}
\definecolor{SAEorange}{rgb}{0.917647058823529,0.443137254901961,0.145098039215686}
\definecolor{SAEred}{rgb}{0.862745098039216,0.16078431372549,0.117647058823529}
\definecolor{SAEblue}{rgb}{0.00392156862745098,0.627450980392157,0.913725490196078}
\definecolor{SAEdblue}{rgb}{0,0.317647058823529,0.584313725490196}
\definecolor{SAEgreen}{rgb}{0.180392156862745,0.694117647058824,0.207843137254902}
\definecolor{SAEdgreen}{rgb}{0,0.466666666666667,0.23921568627451}
\definecolor{SAElgray}{rgb}{0.792156862745098,0.792156862745098,0.784313725490196}
\definecolor{SAEmgray}{rgb}{0.603921568627451,0.607843137254902,0.615686274509804}
\definecolor{SAEdgray}{rgb}{0.380392156862745,0.384313725490196,0.396078431372549}

\usepackage{array,multirow,tabularx,ragged2e}
\newcolumntype{L}[1]{>{\raggedright\let\newline\\\arraybackslash\hspace{0pt}}m{#1}}
\newcolumntype{C}[1]{>{\centering\let\newline\\\arraybackslash\hspace{0pt}}m{#1}}
\newcolumntype{R}[1]{>{\raggedleft\let\newline\\\arraybackslash\hspace{0pt}}m{#1}}

\begin{document}
\begin{frontmatter}

\title{{Physics-based reduced-order modeling of flash-boiling sprays in the context of internal combustion engines}}

%
%
%

\author[mymainaddress]{A. Saha\corref{mycorrespondingauthor}}
\cortext[mycorrespondingauthor]{Corresponding author}
\ead{a.saha@itv.rwth-aachen.de}
\author[mymainaddress]{A. Y. Deshmukh}
\author[mymainaddress,mysecondaryaddress]{T. Grenga}
\author[mymainaddress]{H. Pitsch}

\address[mymainaddress]{Institute for Combustion Technology, RWTH Aachen University, Aachen, Germany}
\address[mysecondaryaddress]{Faculty of Engineering and Physical Science, University of Southampton, Southampton, United Kingdom}

\begin{abstract}
Flash-boiling injection is one of the most effective ways to accomplish improved atomization compared to the high-pressure injection strategy. The tiny droplets formed via flash-boiling lead to fast fuel-air mixing and can subsequently improve combustion performance in engines. Most of the previous studies related to the topic focused on modeling flash-boiling sprays using three-dimensional (3D) computational fluid dynamics (CFD) techniques such as direct numerical simulations (DNS), large-eddy simulations (LES), and Reynolds-averaged Navier-Stokes (RANS) simulations. However, reduced order models can have significant advantages for applications such as the design of experiments, screening novel fuel candidates, and creating digital twins, for instance, because of the lower computational cost. In this study, the previously developed cross-sectionally averaged spray (CAS) model is thus extended for use in simulations of flash-boiling sprays. The present CAS model incorporates several physical submodels in flash-boiling sprays such as those for air entrainment, drag, superheated droplet evaporation, flash-boiling induced breakup, and aerodynamic breakup models. The CAS model is then applied to different fuels to investigate macroscopic spray characteristics such as liquid and vapor penetration lengths under flash-boiling conditions. It is found that the newly developed CAS model captures the trends in global flash-boiling spray characteristics reasonably well for different operating conditions and fuels. Moreover, the CAS model is shown to be faster by up to four orders of magnitude compared with simulations of 3D flash-boiling sprays. The model can be useful for many practical applications as a reduced-order flash-boiling model to perform low-cost computational representations of higher-order complex phenomena.
\end{abstract}


\end{frontmatter}


\begin{abstract}
Flash-boiling injection is one of the most effective ways to accomplish improved atomization compared to the high-pressure injection strategy. The tiny droplets formed via flash-boiling lead to fast fuel-air mixing and can subsequently improve combustion performance in engines. Most of the previous studies related to the topic focused on modeling flash-boiling sprays using three-dimensional (3D) computational fluid dynamics (CFD) techniques such as direct numerical simulations (DNS), large-eddy simulations (LES), and Reynolds-averaged Navier-Stokes (RANS) simulations. However, reduced order models can have significant advantages for applications such as the design of experiments, screening novel fuel candidates, and creating digital twins, for instance, because of the lower computational cost. In this study, the previously developed cross-sectionally averaged spray (CAS) model is thus extended for use in simulations of flash-boiling sprays. The present CAS model incorporates several physical submodels in flash-boiling sprays such as those for air entrainment, drag, superheated droplet evaporation, flash-boiling induced breakup, and aerodynamic breakup models. The CAS model is then applied to different fuels to investigate macroscopic spray characteristics such as liquid and vapor penetration lengths under flash-boiling conditions. It is found that the newly developed CAS model captures the trends in global flash-boiling spray characteristics reasonably well for different operating conditions and fuels. Moreover, the CAS model is shown to be faster by up to four orders of magnitude compared with simulations of 3D flash-boiling sprays. The model can be useful for many practical applications as a reduced-order flash-boiling model to perform low-cost computational representations of higher-order complex phenomena.
\end{abstract}
\newpage
\section{Introduction}
Gasoline direct-injection spark-ignition (DISI) engines have been demonstrated to have the potential for higher thermal efficiencies than port-fuel injection (PFI) engines leading to lower fuel consumption and lower carbon dioxide ($\COtwo$) emissions~\citep{Leach2013}. DISI technology has several other advantages such as improved charge cooling potential, faster transient response, precise fuel metering, and lower cold start emissions~\citep{Reitz2013,Aleiferis2013,Yang2013,Zhao2010}. DISI engines use high-pressure injection systems for atomizing the liquid fuel into small droplets. Although an accurate control of fuel delivery, atomization, and good efficiency with reasonable emissions can be achieved via high-pressure injection systems, there are certain drawbacks associated with this injection mechanism. For example, due to high-pressure injection, the liquid fuel jet exits the injector nozzle normally with higher momentum, and thus increases the chances of spray impingement onto the cylinder liner and/or piston head~\citep{Xu2013}. The formation of a liquid film on the wall surface due to spray impingement will in turn affect the near-wall fuel-air mixing process~\citep{Zhihao2010}, as the fuel film is hard to evaporate and could lead to irregular combustion such as pool fire with subsequent production of soot and unburned hydrocarbons~\citep{Ratcliff2016, Xu2013}. Due to ever-increasing stringent emission regulations, the atomization characteristics of DISI engines also need to be improved further such that the resulting fine spray droplets from enhanced atomization lead to better fuel-air mixing with desired combustion characteristics~\citep{Xu2013}. Increasing injection pressure to extremely high values for achieving superior atomization characteristics has already been shown to be insufficient~\citep{Lefebvre1988}. However, the drawbacks mentioned above with the high-pressure injection strategy can be overcome by using the flash-boiling injection technique~\citep{Sun2021a,Yang2013}. \par 
Flash-boiling injection in DISI engines has become a promising alternative to generate a much finer spray compared to high-pressure injection~\citep{She2010,Schmitz2002,Fujimoto1997,Yamazaki1985}. Injecting liquid fuel into DISI engines operating at part-load, light-load, or idle conditions with early injection strategies for homogeneous-charge engine operation causes explosive vaporization of the fuel jet via bubble nucleation and growth~\citep{Badawy2022,Zeng2012}. This rapid phase-change phenomenon occurs due to the superheating of the liquid fuel upon entering the combustion chamber under the above-mentioned operating conditions. The potentially explosive nature of flash-boiling results in tiny droplets due to the abrupt disintegration of the liquid jet, which in turn enhances the mixture homogeneity between air and fuel by increasing the vaporization rate~\citep{Price2018}, widening the spray plume due to the increased radial expansion via bubble growth, and reducing the droplet velocities, thus leading to shorter penetration~\citep{Badawy2022}. \par
Flash-boiling consists of different sub-processes such as nucleation, bubble growth, and droplet burst~\citep{Senda1994}. The small length and time scales associated with these processes make it difficult to accurately quantify their influences on flash-boiling sprays~\citep{Dietzel2020,Price2018}. Many researchers have studied the flash-boiling phenomena at a microscopic single droplet level to have an in-depth understanding of the above-listed subprocesses (such as \cite{Saha2023,Saha2022,Saha2021,Wang2020,Xi2017,Li2017}, among the most recent). \cite{Brown1962} were the first to investigate the influence of flash-boiling at a macroscopic scale on the breakup and atomization process of superheated water and freon-11 jets. \cite{Sher1977} quantified the flash-boiling spray formed by the binary mixture of toluene and freon 22 from pressure cans in terms of droplet size distributions for different pressures and temperatures. \cite{Adachi1997} reported a significant increase in fuel vapor concentration and subsequently more homogeneous fuel-air mixture formation from their experimental and theoretical investigations on the atomization process of superheated \pen~sprays. \cite{Vanderwege1998} studied the effect of fuel volatility on sprays from high-pressure swirl injectors for fuel mixtures of doped and undoped \isooct~and indolene, and reported a 40$\%$ reduction in droplet diameter under flash-boiling conditions. \cite{Kale2019} investigated the flash-boiling behavior of alcohol fuels using a direct-injection (DI) injector under engine-like hot injector body conditions. They also reported a significant reduction in droplet diameter (58.45$\%$ and 54.5$\%$ for butanol and \isobut, respectively) at elevated fuel temperatures due to the occurrence of flash-boiling. \cite{Senda2008} performed experiments with fuel mixtures of \trid~and liquefied $\COtwo$ to investigate the flash-boiling spray combustion characteristics. They found that the flash-boiling of the fuel mixture leads to a significant reduction in soot and  $\NOx$ emissions. The brake-specific fuel consumption was also observed to decrease due to the formation of an advanced flammable mixture resulting from flash-boiling. \cite{Aleiferis2010a} and \cite{Serras2010} studied the combined effect of cavitation and flash-boiling on spray characteristics of different fuels using an optical DI injector nozzle. They revealed that an increase in the cavitation phenomenon inside the nozzle hole at higher fuel temperatures results in a large number of vapor bubbles, which then act as a strong source of nucleation sites and lead to an increase in the rate of superheated fuel vaporization. Several other experimental studies on flash-boiling injection are available in the literature (such as \cite{Sun2021b,Sun2020,Guo2018,Guo2017a,Mojtabi2014,Aleiferis2010b}, to name a few), which confirm the potential of flash-boiling sprays in producing superior spray atomization along with improved combustion characteristics in DISI engines. \par
However, the characteristics of flash-boiling sprays deteriorate depending on the degree of superheating and the number and vicinity of the nozzle holes. With increasing superheating degrees, the radial expansion of the flash-boiling spray plumes increases, which may increase the possibility of jet-to-jet interactions and consequently the collapsing of the spray plumes. The collapsed flash-boiling sprays are associated with higher momentum flux and may lead to an increase in piston wall-wetting due to spray impingement~\citep{Duronio2021,Li2019,Li2018a} and lubricant dilution, which are known to be one of the main sources of super-knock and engine damage~\citep{Wang2017c}.  
Studies on the collapsed flash-boiling spray characteristics and the detailed mechanism behind the spray collapse can be found in the literature (such as \cite{Zhou2018,Li2018b,Wang2017a,Wang2017b,Guo2017b}, to name a few).  \par 
Significant efforts were also made in developing numerical methods for modeling flash-boiling spray characteristics. 
\cite{Zuo2000} presented a superheated spray and vaporization model for studying the evolution and vaporization behavior of flashing sprays in GDI engines. \cite{Zeng2001} developed an atomization model for flash-boiling sprays and concluded that the combined effect of aerodynamic forces and bubble expansion is responsible for breakup under flash-boiling conditions. \cite{Kawano2004} integrated the bubble nucleation, growth, and disruption sub-models into the KIVA3V code, and  numerically investigated the flash-boiling characteristics of multicomponent fuels. \cite{Price2016} proposed a numerical framework for modeling flash-boiling sprays using a Lagrangian particle tracking (LPT) technique. The spray collapse and recirculations of droplets were well predicted by their model in comparison with the experimental measurements. \cite{Price2018} later applied this model to investigate the flash-boiling spray characteristics of high-volatility fuels (such as \pen, \hexane, \isohex,~and ethanol) as well as low-volatility fuels (such as \isooct~and \but) over a wide range of injection systems. \cite{Guo2019b} numerically investigated the flashing \textit{n}-hexane sprays using the homogeneous relaxation model (HRM) in a diffuse Eulerian framework. They observed an under-expanded flashing jet due to the explosive evaporation within the intact liquid core, which was found to exist in the near nozzle regime. With increasing superheating degrees, the shock-wave structure, known as `Mach-disk', was also identified at some distance from the nozzle exit. \cite{Duronio2021} developed a Lagrangian flash-boiling breakup model in OpenFOAM and studied the spray characteristics of the Engine Combustion Network (ECN) Spray G injector under flash boiling conditions for different injection pressures. \cite{Duronio2022} later simulated the internal nozzle flow in an Eulerian framework using the HRM model and coupled it with their previously developed Lagrangian external spray simulations to investigate the effect of in-nozzle phase change on global spray characteristics. \mytab{SummaryTab1} and \mytab{SummaryTab2} provide a comprehensive summary of the experimental and computational analysis contents of flash-boiling sprays, respectively, from several preceding studies mentioned earlier. \par
Although the three-dimensional (3D) computational fluid dynamics (CFD) simulations are necessary to get a detailed fundamental understanding of the underlying physics of the flash-boiling phenomena, the high computational cost associated with these simulations makes their application in the design of experiments, system simulations, creating digital twins, and screening of novel fuel candidates difficult. Model order reduction from 3D to 1D or 0D while preserving the essential multiphysics information is therefore useful as long as sufficient accuracy can be achieved. For example, in the fuel design process, each and every fuel candidate needs to undergo an extensive testing process before being used in real combustion systems. Assessing the spray characteristics and combustion performance is one of the crucial steps in this testing phase. However, conducting experiments on an engine test bench to investigate the spray and combustion characteristics for a large number of fuel candidates is extremely difficult~\citep{Deshmukh2022}. 
The physics-based reduced-order models (ROM) can also be used as a digital twin of the internal combustion engine for closed-loop feedback control of the combustion performance~\citep{Deshmukh2022}. Moreover, the simplicity of these ROMs allows for their interactive utilization with multi-dimensional spray simulations, as previously demonstrated by \cite{Wan1997}.\par       
\cite{Sazhin2001} proposed simplified analytical expressions for obtaining the spray tip penetration during the initial stages and the later two-phase flow regimes. \cite{Desantes2006} derived a theoretical model to investigate the influence of the injection parameters on the macroscopic spray characteristics based on the assumption of the conservation of the momentum flux along the spray axis. \cite{Pastor2008} investigated the global spray characteristics of the transient inert diesel sprays using a 1D Eulerian spray model considering the mixing-controlled processes and locally-homogeneous flow field. Later, \cite{Desantes2009} extended it to reactive diesel spray cases. However, the droplet dynamics were neglected in these 1D Eulerian spray models. \cite{Wan1997} was the first to derive a 1D cross-sectionally averaged spray model (CAS) from 3D multiphase governing equations for diesel sprays in the context of compression-ignition (CI) engines considering droplet dynamics. Recently, \cite{Deshmukh2022} proposed some crucial improvements to the original CAS model by incorporating an additional vapor transport equation and state-of-the-art droplet breakup and evaporation models and found that the CAS model is able to predict the trend in inert subcooled spray characteristics reasonably well compared to the original work by \cite{Wan1997} for different fuels under a wide range of operating conditions. They also extended the CAS model later to reactive turbulent spray cases~\citep{Deshmukh2022b}. Although considerable efforts have been made to the development of the ROMs for subcooled sprays, studies on the 1D physics-based ROM development for macroscopic characterization of the flash-boiling sprays are still scarce in the literature. Most computational studies on flash-boiling sprays are based on a 3D CFD approach. In this work, the CAS model developed by \cite{Deshmukh2022} is further extended for the simulation of flash-boiling sprays. The extended CAS model is applied to different fuels under engine-like conditions, which are susceptible to flash-boiling. It incorporates several important physical sub-processes in flash-boiling sprays including air entrainment, drag, droplet internal as well as external vaporization, droplet heating, flash-boiling induced breakup, and aerodynamic breakup models. The proposed physics-based ROM for flash-boiling sprays is found to capture reasonably the trends in spray penetration lengths for different fuels under flash-boiling conditions. \par   
The remaining manuscript is structured as follows. \mysec{CAS_model} introduces the new extended CAS model in detail. The numerical methods and the solution procedure are briefly discussed in \mysec{Num_Method}. The detailed description of the experimental cases used for model validation is presented in \mysec{Case_Des}. In \mysec{results}, the macroscopic spray characteristics predicted by the previous CAS model as well as the newly developed CAS model are discussed. Finally, the findings of the present study are summarized in \mysec{concl}. 
\renewcommand{\arraystretch}{1.0}

\newcolumntype{L}[1]{>{\raggedright\arraybackslash}p{#1}}
 
\newcolumntype{C}[1]{>{\centering\arraybackslash}p{#1}}
 
\newcolumntype{R}[1]{>{\centering\arraybackslash}p{#1}}
\begin{table}[!t]
\begin{center}
\begin{tabular}{L{1cm}C{4cm}R{7.8cm}}
\toprule
\textbf{Year} & \textbf{Authors} & \textbf{Experiment details} \\
\midrule
1962 & \cite{Brown1962} & {Exploration of the influence of flash-boiling at macroscopic scale on the atomization processes} \\
1977 & \cite{Sher1977} & Quantification of droplet size distributions over a wide range of flash-boiling conditions \\
1997 & \cite{Adachi1997} & Fuel vapor concentration characterization in \pen~flash-boiling sprays \\
1998 & \cite{Vanderwege1998} & Study of the effect of fuel volatility on flash-boiling sprays in high-pressure swirl injectors \\
2002 & \cite{Schmitz2002} & Investigation of the impact of injector temperature on DI gasoline engine flash-boiling sprays \\
2008 & \cite{Senda2008} & Characterization of flash-boiling spray combustion \\
2010 & \cite{Aleiferis2010a} & Study of the influence of cavitation on flash-boiling behavior of hydrocarbon and alcohol fuels\\
2010 & \cite{She2010} &   Impact of high-temperature flash-boiling on homogeneous-charge compression ignition diesel engine performance\\
2012 & \cite{Zeng2012} & Multi-hole injector flash-boiling spray characteristics for alcohol fuels \\
2017 & \cite{Li2017} & Quantitative investigation of the droplets morphology variation and breakup process in flash-boiling ethanol fuels\\
2019& \cite{Kale2019} & Study of alcohol fuels flash-boiling behavior using DI injector \\
2020 & \cite{Wang2020} & Investigation of bubble nucleation and micro-explosion phenomena in superheated jatropha oil droplets\\
2022 & \cite{Badawy2022} & Effect of fuel temperature, ambient pressure, and fuel properties on multi-hole injector flash-boiling sprays\\

\bottomrule
\end{tabular}
\end{center}
\caption{Previous experimental studies on flash-boiling single droplets and sprays.}
\label{SummaryTab1}
\end{table}

\renewcommand{\arraystretch}{1.0}

\newcolumntype{L}[1]{>{\raggedright\arraybackslash}p{#1}}
 
\newcolumntype{C}[1]{>{\centering\arraybackslash}p{#1}}
 
\newcolumntype{R}[1]{>{\centering\arraybackslash}p{#1}}
\begin{table}[!t]
\begin{center}
\begin{tabular}{L{1cm}C{4cm}R{8cm}}
\toprule
\textbf{Year} & \textbf{Authors} & \textbf{Simulation details}\\
\midrule
1994 & \cite{Senda1994} & Analytical modeling of atomization and evaporation processes in flash-boiling sprays \\
2000 & \cite{Zuo2000} & Comprehensive modeling of superheated vaporization and breakup processes in GDI engines \\
2001 & \cite{Zeng2001} &  Modeling of atomization process in flash-boiling sprays \\
2004 & \cite{Kawano2004} & Modeling of atomization and vaporization processes in multi-component flash-boiling sprays \\
2017 & \cite{Xi2017} & Lagrangian modeling of single droplet flash-boiling  \\
2018 & \cite{Price2018} & Lagrangian modeling of multi-hole flash-boiling spray over a broad range of injection systems and operating conditions \\
2017 & \cite{Guo2019b} &  Eulerian modeling of flash-boiling characteristics of \textit{n}-hexane fuel \\
2021 & \cite{Duronio2021} & Lagrangian modeling of \isooct~flash-boiling behavior in ECN Spray G injector \\
2022 & \cite{Duronio2022} & Investigation of the influence of in-nozzle phase change on flash-boiling spray using a combined Eulerian-Lagrangian framework \\
2020 & \cite{Dietzel2020} & Modeling of the influence of bubble interactions in flash-boiling cryogenic liquids using DNS \\
2022 & \cite{Saha2022} & Modeling of single droplet flash-boiling behavior of e-fuels considering internal and external vaporization process\\
2023 & \cite{Saha2023} & Reduced-order modeling of the influence of bubble interactions in highly volatile e-fuel microdroplets\\
\bottomrule
\end{tabular}
\end{center}
\caption{Previous simulation studies on flash-boiling single droplets and sprays.}
\label{SummaryTab2}
\end{table}
\section{Cross-sectionally averaged spray (CAS) model}\label{CAS_model}
\begin{figure}
\centering
\includegraphics[width=370pt]{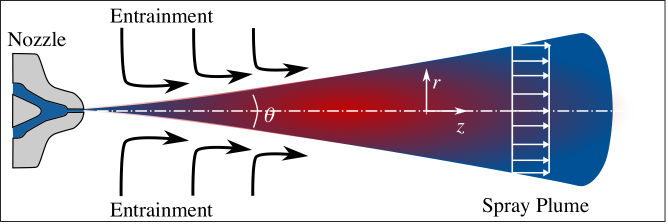}   
\caption{Schematic of the CAS model.}
\label{CAFS}
\end{figure}

The 3D multiphase governing equations for a complete spray~\citep{Hiroyasu1980} are reduced to 2D by assuming azimuthal symmetry and then radially integrated to obtain a one-dimensional system of equations. The reader is referred to \cite{Wan1997} for more details about the model reduction. The governing equations (GEs) for the newly developed CAS model are given by 
\begin{linenomath*}
\begin{align}
		\Dg( \rhohat \Yhata b^2) & = \wenta b \label{eq:amb_cont}, \\
		\Dg( \rhohat \Yhatv b^2) & =\bigmean{\wvapextotal+\wvapswl} b^2 \label{eq:vap_cont}, \\
		\Dg((\rhohat\Yhatg \uhatg b^2) & = -\bigmean{\wdrag} b^2 + \bigmean{\wvapextotal+\wvapswl} \uhatl  b^2 \label{eq:gas_mom}, \\
  \Dg(\rhohat\Yhatg \Thatg b^2) & =   -\bigmean{\wheat}\frac{\Cl}{\Cpg}b^2 + \bigmean{\left(\wvapextotal+\wvapswl\right)}\frac{\Cpv}{\Cpg}\Thatv b^2+\wenta\frac{\Cpa}{\Cpg}\Thata b\label{eq:gas_energy},\\
		\Dl((\rhohat\Yhatl b^2) & =  - \bigmean{\wvapextotal+\wvapswl} b^2 \label{eq:liq_cont}, \\
		\Dl((\rhohat\Yhatl \uhatl b^2) & = \bigmean{\wdrag}b^2 - \bigmean{\wvapextotal+\wvapswl} \uhatl  b^2 \label{eq:liq_mom}, \\
		\Dl(\rhohat\Yhatl \mean{\ddhat} b^2) & = -\bigmean{\frac{\wbre}{2\ddhat}} b^2 - \frac{4}{3}\bigmean{\wvapextotal\ddhat} b^2- \frac{2}{3}\bigmean{\wvapswl\ddhat} b^2 \label{eq:d_trans}, \\
		\Dl(\rhohat\Yhatl \Thatd b^2) & =   \bigmean{\wheat} b^2 - \bigmean{\left(\wvapextotal+\wvapswl\right)}\Thatd b^2 \label{eq:liq_energy},
\end{align}
\end{linenomath*}
where $\rho$ denotes the density, $b(z,t)$ the spray width, $z$ the axial coordinate, $t$ the temporal coordinate, $Y$ the mass fraction, $u$ the velocity, $d$ the droplet diameter, $C$ the specific heat capacity, $C_\text{p}$ the specific heat capacity at constant pressure, and $T$ the temperature. The subscripts `d', `l', `g', `a', and `v' refer to the droplet variables, liquid phase, gas phase, ambient gas, and vapor, respectively. The radially integrated differential operator, $\mathrm{D}_{\text {i}}(\bar{\rho} \hat{\phi} b^2)$, is defined as
\begin{equation}
    \mathrm{D}_{\text {i}}(\bar{\rho} \hat{\phi} b^2)=\frac{\partial}{\partial t}(\bar{\rho} \hat{\phi} b^2)+\frac{\partial}{\partial z}(\bar{\rho} \hat{\phi} \hat{u}_{\mathrm{i}} b^2) \hskip0.5cm\text { with}\hskip0.5cm i=\mathrm{g,\,l} \text {,}
\end{equation}
where $\phi$ represents the quantity of interest. The density-weighted cross-sectional average of $\phi$ is defined as
\begin{equation}
    \bar{\rho} \hat{\phi} b^2=2 \int_0^{\infty} \rho \phi r \mathrm{~d} r,
\end{equation}
where $r$ is the radial coordinate. The cross-sectional averaging with $\phi=1$ and density-weighted cross-sectional averaging are denoted by the `overline' ($\bar{\cdot}$) and `hat' ($\hat{\cdot}$) operators, respectively. The operator, $\mean{\cdot}$, in the GEs represents the expectation value of any term, $\zeta(\ddhat)$, which is a function of droplet diameter and is defined as 
\begin{equation}
    \langle\zeta\rangle=\int \zeta(\hat{d}^{\prime}) \mathcal{P}(\hat{d}^{\prime}) \mathrm{d} \hat{d}^{\prime},
\end{equation}
where $\mathcal{P}(\hat{d}^{\prime})$ denotes the droplet size distribution. However, due to the difficulties associated with the use of polydisperse droplet size distribution~\citep{Deshmukh2022}, the droplets are assumed to be monodisperse in this study. A schematic of the CAS model is shown in \myfig{CAFS}. The source terms on the right-hand side of the GEs describe different sub-models for the physical processes, such as air-entrainment ($\wenta$), drag ($\wdrag$), heat transfer ($\wheat$), droplet breakup ($\wbre$), vaporization ($\wvapextotal$), and droplet expansion ($\wvapswl$) in flash-boiling sprays. The details of the sub-models are discussed in the next subsections.
\subsection{Air entrainment model}\label{entrainment}
Air entrainment into the spray plume describes the air mass flow across the spray boundary. The spray morphology is significantly influenced by the entrained air \citep{Ali2022}. The higher the entrainment rate, the better will be the mixing between the injected fuel and ambient air, and subsequently, the combustion performance will be improved. The air entrainment source term is modeled as~\citep{Deshmukh2022}
\begin{equation}
    \dot{\omega}_{\mathrm{ent}, \mathrm{a}}=\rho_{\mathrm{a}} \beta \hat{u}_{\mathrm{g}},
\end{equation}
where $\beta$ is the spreading coefficient defined as $\beta=\tan (\theta/ 2)$ and $\theta$ (in degree) the spray cone angle. \subsubsection{Spray cone angle model}
The spray cone angle, which was previously modeled using the \cite{Hiroyasu1980} correlation, is updated in this work for superheated conditions as \citep{Price2018} 
\begin{equation}\label{Cone}
    \theta=\log \left(\frac{R_p^2 \Theta^3}{m_\mathrm{a}^2}\right),
\end{equation}
where $\Rp$ denotes the ratio of saturated vapor pressure ($\Pv$) corresponding to the injection temperature ($\Tinj$) and ambient gas pressure ($\Pg$), and $m_\mathrm{a}$ the atomic mass of the injected fuel. $\Theta$ is the nondimensional surface tension defined as
\begin{equation}
    \Theta=\frac{a \sigma}{k_\mathrm{b}\Tinj} \hskip0.5cm\text { with}\hskip0.5cm a=\left({36\pi}{\vl^2}\right)^{1/3}\text {,}
\end{equation}
where $k_\mathrm{b}$ is the Boltzmann constant, $a$ the molecular surface area, $\vl$ the liquid molecular volume, and $\sigma$ the surface tension. 
\subsection{Drag model}
The steady-state drag force on a droplet can be expressed as \citep{Crowe2012}
\begin{equation}
    F_\mathrm{ss,drag}=\frac{1}{2}\rhog\Cdrag\Ad\left(\ug-\ul\right)\left|\ug-\ul\right|,
\end{equation}
where $\Ad$ is the droplet surface area and $\Cdrag$ the droplet drag coefficient computed as~\citep{Wallis1969}
\begin{equation}
    \Cdrag=\left\{\begin{array}{ll}
\frac{24}{\Red}(1 + \frac{1}{6}\Red^{2/3}) & \quad \text{for $\Red \leq 1000$ } \\
0.424 & \quad \text{for $\Red > 1000$}.
\end{array}\right.
\end{equation}
$\Red$ denotes the droplet Reynolds number and is given by
\begin{equation}
    \Red = \rhog\left|\ug-\ul\right|\ddhat/\mug,
\end{equation}
where $\mug$ is the gas-phase molecular viscosity. The source term due to the steady-state drag force on a droplet of diameter $\ddhat$ is thus computed as
\begin{equation}
    \wdrag = \frac{3 \Cdrag \rhog \rhohat \Yhatl(\uhatg-\uhatl)|\uhatg-\uhatl|}{4\rhol \ddhat}.
	\label{eq:total_drag}
\end{equation}
\subsection{Superheated droplet breakup model} 
\cite{Deshmukh2022} modeled the droplet breakup only via the aerodynamic breakup mechanism as the sub-cooled liquid droplet atomization is known to be mainly controlled by the aerodynamic forces acting on the droplet surface. However, for the flash-boiling droplet, the superheating degree plays an important role in determining the atomization characteristics. The spontaneous growth of the vapor bubbles and the aerodynamic forces compete with each other during the flash-boiling spray atomization process. Micro-explosions due to the bubble growth dominate in the regime of high superheating degrees, whereas aerodynamic forces dominate in low superheating degree regimes~\citep{Zeng2001}. Thus, in this work, the CAS model has been integrated with a hybrid breakup model which includes both the thermal and the aerodynamic breakup mechanisms. \par 
A liquid droplet generates multiple child droplets at the end of the breakup. However, a continuous thermal and aerodynamic breakup approach is considered in the CAS model for simplicity, which results in a continuous reduction of the droplet diameter rather than generating multiple child droplets~\citep{Deshmukh2022}. The source term due to the droplet breakup is expressed as $\wbre = \Kbre \rhohat$, where $\Kbre$ is the breakup coefficient modeled either via $\Kbrethm$ or via $\Kbreaerod$ depending on the breakup length and time scales. $\Kbrethm$ and $\Kbreaerod$ describe the breakup coefficients resulting from the micro-explosion (herein referred to as `thermal breakup') and aerodynamic force-induced breakup (herein referred to as `aerodynamic breakup') mechanism, respectively. Details on the calculation of the breakup coefficients are discussed in the following subsections. \par
\subsubsection{Thermal breakup}
The breakup coefficient of the thermal breakup is modeled as
\begin{equation}
    \Kbrethm = \frac{2\ddhat (\ddhat-\dstthm)}{\taubthm}\hskip0.5cm\text{with}\hskip0.5cm\taubthm=A_0\frac{\ddhat}{\eta_\mathrm{thm}\Omega_\mathrm{thm}}\hskip0.4cm\text{and}\hskip0.4cm\dstthm=A_1\eta_\text{thm},
	\label{eq:kbrethm}
\end{equation}
where $\dstthm$ and $\taubthm$ denote the stable droplet diameter and breakup time associated with the thermal breakup mechanism, respectively. The amplitude of the disturbance due to the bubble growth ($\eta_\text{thm}$) and the disturbance growth rate ($\Omega_\mathrm{thm}$) are defined as 
\begin{equation}\label{etathm}
    \eta_\mathrm{thm}=\left(\frac{6\epcrit\Vdrop}{\pi\Nbub\left(1-\epcrit\right)}\right)^{1/3}\hskip0.5cm\text{and}\hskip0.5cm\Omega_\mathrm{thm}=\frac{2\Rdot}{\Do}\hskip0.5cm\text{with}\hskip0.5cm\Rdot=\sqrt{\frac{2}{3}\frac{\Pv\left(\Thatd\right)-\Pg}{\rhol}},
\end{equation}
where $\epcrit$ denotes the critical void fraction given by $\epcrit=V_\mathrm{b}/\left(V_\mathrm{b}+\Vdrop\right)$. $V_\mathrm{b}$ and $\Vdrop$ are the total volume of the vapor bubbles and liquid in the superheated droplet, respectively. In the CAS model, a value of 0.55 is considered for $\epcrit$~\citep{Kawano2004}. $\Nbub$ represents the total number of vapor bubbles in the superheated droplet and is  calculated using the empirical bubble number density proposed by \cite{Senda1994} as 
\begin{equation}
    n=5.757 \times 10^{12} \cdot \exp \left(\frac{-5.279 \mathrm{~K}}{\DeltaT}\right)\hskip0.5cm\text{with}\hskip0.5cm\DeltaT = \Tinj-\Tb(\Pg),
\end{equation}
where $\DeltaT$ is the superheating degree of the liquid droplet.
\subsubsection{Aerodynamic breakup}
A combined Kelvin–Helmholtz (KH) - Rayleigh–Taylor (RT) breakup
model is incorporated in the CAS model for the aerodynamic breakup. The aerodynamic breakup coefficient is expressed as
\begin{equation}
    \Kbreaerod = \frac{2\ddhat (\ddhat-\dstaero)}{\taubaero}.
	\label{eq:kbreaero}
\end{equation}
The stable droplet diameter ($\dstaero$) and breakup time ($\taubaero$) associated with the aerodynamic breakup are computed either via the KH model as
\begin{equation}
    	\dstKH = 2 B_0 \Lambda_{\mathrm{KH}},\hskip0.5cm
		\taubKH  = 3.788 B_1 \frac{\ddhat}{2 \Lambda_{\mathrm{KH}} \Omega_{\mathrm{KH}}}
\end{equation}
or via the RT model as
\begin{equation}
		\dstRT  = C_3 \Lambda_{\mathrm{RT}},\hskip0.5cm
		\taubRT  = \Omega_{\mathrm{RT}}^{-1},
\end{equation}
where $\Lambda$ is the wavelength of the fastest-growing wave and $B_0$ the model constant equal to 0.61. The detailed calculation of $\Lambda$ and $\Omega$ corresponding to the KH-RT breakup model can be found in \cite{Patterson1998}.\par
The breakup is modeled via the thermal breakup mechanism if $\ddhat>\dstthm$ and $\taubthm<\taubaero$, otherwise, the breakup takes place via the aerodynamic breakup mechanism. The RT model is used for aerodynamic breakup if $\ddhat>\dstRT$ and $\taubRT<\taubKH$, else the KH model is considered. The breakup model constants, $A_0$, $A_1$, $B_1$, and $C_3$, depend on the injector nozzle geometry and therefore need to be tuned for a given injector to match the experimental liquid length~\citep{Deshmukh2022}. It is to be noted that for a given injector, once the tuning is performed, the breakup model can be used for any fuel under any operating conditions without further tuning the model constants. 
\subsection{Superheated droplet vaporization model}
\cite{Deshmukh2022} previously modeled evaporation using the \cite{Miller1999} model, which is not valid for superheated droplets. The phase transition in superheated liquid droplets occurs in two distinct ways: (1) by spontaneous nucleation and subsequent growth of the vapor bubbles in the droplet, herein referred to as `internal vaporization', and (2) by vaporization from the droplet's external surface due to the internal as well as external temperature gradient, herein referred to as `external vaporization'~\citep{Yang2017, Saha2022}. The evaporation model in the CAS formulation is thus updated with the above-mentioned vaporization phenomena.  
\subsubsection{Internal vaporization}
\begin{figure}[h]
\centering
\includegraphics[width=370pt]{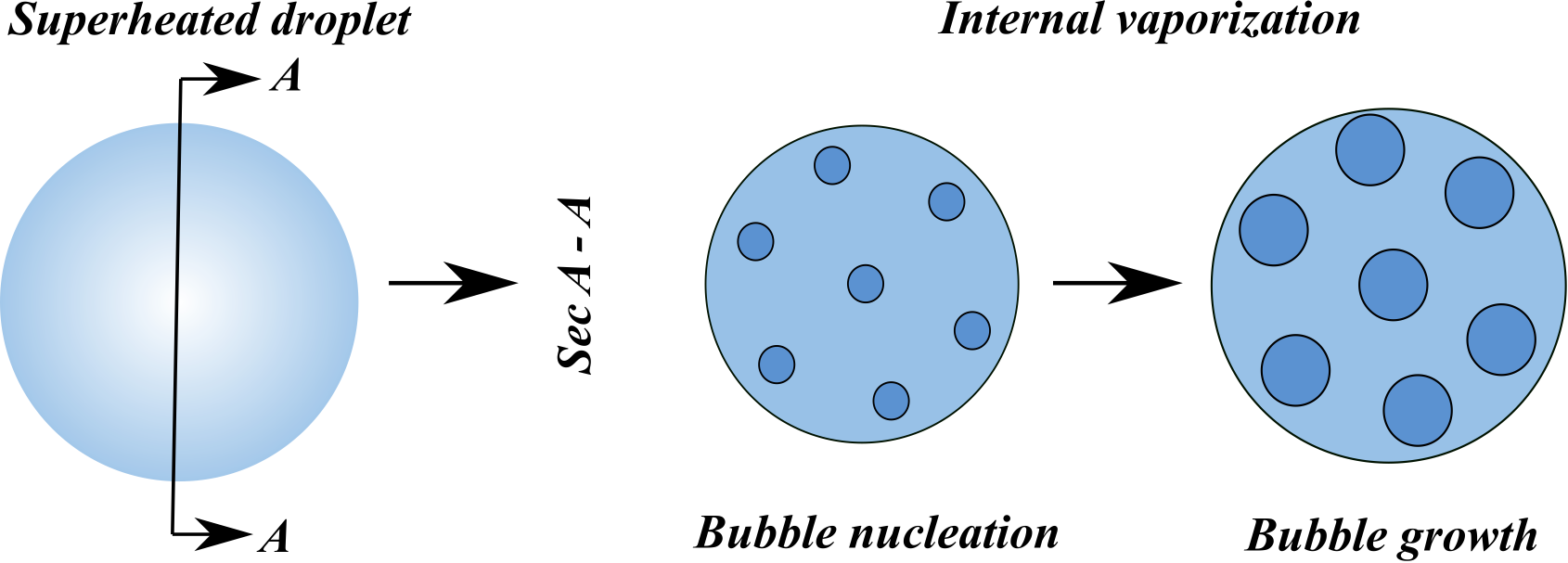}   
\caption{Schematic of the internal vaporization process~\citep{Saha2023}.}
\label{intvap}
\end{figure}

The internal vaporization via the formation of vapor bubbles causes the droplet to expand in the radially outward direction in the superheated regime. A schematic of the internal vaporization process is shown in \myfig{intvap}. The source term of droplet expansion is modeled as
\begin{equation}
    \wvapswl=\frac{3\Kvapswl\rhohat\Yhatl}{2\ddhatsq},
	\label{eq:kvapswl}
\end{equation}
where $\Kvapswl$ is expressed as~\citep{Saha2023} 
\begin{equation}
    \Kvapswl=\frac{4\ddhat \rho_v\Rdot}{\rho_l}N_\text{bub}.
\end{equation}
\subsubsection{External vaporization}
The superheated liquid droplet undergoes a phase change from its external surface due to the heat transfer from the inner core of the droplet. The external phase transition may also take place due to the temperature gradient between the droplet outer interface and the ambient gas. The source term for the vaporization due to the heat flux from the droplet's inner core is modeled as $\wvapint={3\Kvapint\rhohat\Yhatl}/({2\ddhatsq})$. The vaporization coefficient, $\Kvapint$, is given by~\citep{Adachi1997} 
\begin{equation}
    \Kvapint=\frac{4 \ddhat h_\mathrm{1}(\Thatd-\Tb)}{\rho_lL\left(\Tb\right)},
\end{equation}
where $\Tb$ denotes the droplet boiling temperature and $L$ is the latent heat of vaporization at $\Tb$. The internal heat transfer coefficient, $h_\mathrm{1}$ (in kW/m$^2$K), is approximated as \citep{Adachi1997}
\begin{equation}
    h_\mathrm{1}=\left\{\begin{array}{ll}
0.76(\Thatd-T_{\mathrm{b}})^{0.26} & (0<\Thatd-T_{\mathrm{b}}<5) \\
0.027(\Thatd-T_{\mathrm{b}})^{2.33} & (5<\Thatd-T_{\mathrm{b}}<25) \\
13.8(\Thatd-T_{\mathrm{b}})^{0.39} & (\Thatd-T_{\mathrm{b}}>25).
\end{array}\right.
\end{equation}
The vaporization source term due to the temperature gradient between the droplet's external surface and the surrounding gas is modeled as $\wvapext={3\Kvapext\rhohat\Yhatl}/({2\ddhatsq})$, where the vaporization coefficient, $\Kvapext$, is given by~\citep{Adachi1997}
\begin{equation}\label{extVap}
      \Kvapext=\frac{4 \ddhat f_3h_\mathrm{2}|\Thatg-\Thats|}{\rho_lL(\Thats)},
\end{equation}
where $h_\mathrm{2}$ represents the external heat transfer coefficient and $\Thats$ the droplet surface temperature, which is assumed to be equal to $\Tb$ in the superheated regime \citep{Saha2022}. The additional Stefan flow caused by the mass flux due to the heat transfer from the droplet's inner core counteracts the external temperature gradient-based vaporization process and may significantly reduce the heat flux to/from the droplet's outer surface~\citep{Zuo2000}. The factor $f_3$ is an evaporative heat transfer correction factor, which is introduced here to take into account the above-mentioned phenomena, and is defined as
\begin{equation}
    f_3=\frac{1}{1-\text{PR}}\hskip0.5cm\text{with}\hskip0.5cm\text{PR}=\frac{\Pg}{\Pv(\Thatd)}.
\end{equation} 
Combining the contributions from the different external vaporization sources, such as $\wvapint$ and $\wvapext$, yields the source term due to the total external vaporization in the superheated regime as
\begin{equation}
    \wvapextotal = \frac{3\Kvapsup\rhohat\Yhatl}{2\ddhatsq}\hskip0.5cm\text{with}\hskip0.4cm\Kvapsup=\Kvapint+\Kvapext.
\end{equation}
When the superheated liquid droplet cools down to a temperature below $\Tb$, the standard non-equilibrium evaporation model proposed by \cite{Miller1999} is used to model the vaporization process. The vaporization coefficient in this sub-cooled regime is given by~\citep{Deshmukh2022}
\begin{equation}
    \Kvapsub = 4 \frac{\rhog \Gvg}{\rhol}\;\mathrm{ln} (1+\BMd) \Shd,
	\label{eq:kvap}
\end{equation}
where $\Gvg$ denotes the diffusion coefficient of fuel vapor in the ambient gas mixture and $\BMd$ is the Spalding mass transfer number. The Sherwood number, $\Shd$, is computed using the \cite{Ranz1952} correlation as 
\begin{equation}
    \Shd = 2 + 0.552~\Red^{1/2}~\Scg^{1/3},
	\label{eq:sherwoodno_miller}\\
\end{equation}
where the gas-phase Schmidt number, $\Scg$, is defined as 
\begin{equation}
\Scg = \frac{\mug\left(\Tref\right)}{\rhog\left(\Tref\right) \Gvg\left(\Tref\right)}.
\end{equation}
For additional details on the subcooled vaporization modeling, the reader is referred to \cite{Deshmukh2022}. In the CAS model, the liquid properties such as density ($\rhol$), viscosity ($\mul$), and surface tension ($\sigma$) are assumed to be constant throughout the liquid phase and computed at the injection temperature  ($\Tinj$). The \cite{Wilke1950} formula is used to evaluate the gas-phase mixture properties such as viscosity ($\mug$) and thermal conductivity ($\lambdag$), whereas the mixture-specific heat capacity at constant pressure ($\Cpg$) is evaluated using the linear mixing rule~\citep{Deshmukh2022}. All the gas-phase mixture properties and the correlations in the CAS model are computed at reference temperature ($\Tref$) obtained by the one-third rule \citep{Hubbard1975}. The derivation of the vaporization coefficients is provided in \ref{App_evap}.
\subsection{Gas-phase energy transport}
\cite{Deshmukh2022} calculated the gas phase temperature, $\Thatg$, by assuming a homogeneous mixture of fuel vapor and ambient gas. In the present work, a more accurate formulation has been incorporated for obtaining $\Thatg$ considering the following physical phenomena: (1) the energy dissipation/deposition due to the heat transfer between the droplets and the gas-phase denoted by ${\wheat}(\Cl/\Cpg)b^2$, (2) the energy transport by the vaporized fuel into the gas phase expressed as ${\left(\wvapextotal+\wvapswl\right)}(\Cpv/\Cpg)\Thatv b^2$, and (3) the energy transport by the fresh ambient air into the spray plume due to the air-entrainment phenomenon given by $\wenta(\Cpa/\Cpg)\Thata b$. The fuel vapor temperature, $\Thatv$, in \myeq{eq:gas_energy} is calculated based on the energy balance between the liquid and the fuel vapor as
\begin{equation}
    \Thatv=\frac{\Cl\Thatd-L(\Thats)}{\Cpv}.
\end{equation}
\subsection{Heat transfer model}
The droplet bulk temperature in the CAS model was calculated using the infinite-conductivity model proposed by \cite{Miller1999} only considering subcooled heat transfer. In this work, the energy balance equation has been updated for the superheated droplet as
\begin{equation}\label{HeatTransfer1}
    \frac{\mathrm{d}\Thatd}{\mathrm{d}t}=\underbrace{\frac{6f_\text{2,sup} \Nud\lambdag(\Tref) (\Thatg - \Thats)}{\rhol \ddhatsq \Cl(\Thatd)}}_\text{conductive heat flow} - \underbrace{\overbrace{\frac{3 \Kvapsup L(\Thats)}{2 \ddhat^2 \Cl(\Thatd)}}^\text{external vaporization}- \overbrace{\frac{3 \Kvapswl L(\Thatd)}{2 \ddhat^2 \Cl(\Thatd)}}^\text{internal vaporization}}_\text{evaporative cooling}=\Kheatsup.
\end{equation}
The derivation of the heat transfer coefficient, $\Kheatsup$, is given in \ref{App_heat}. In \myeq{HeatTransfer1}, the first term on the right-hand side describes the conductive heat flow per unit time between the droplet surface and the external ambient. The second and third terms represent the evaporative cooling of the droplet due to external and internal vaporization, respectively. $f_2$ is the analytical evaporative heat transfer correction factor defined as 
\begin{equation}
	f_\text{2,sup} = \frac{\xi_\text{sup}}{e^{\xi_\text{sup}}-1}.
	\label{eq:f2}
\end{equation}
$\Nud$ denotes the Nusselt number and is computed using \cite{Ranz1952} correlations as
\begin{equation}
    \Nud = 2 + 0.552~\Red^{1/2}~\Prg^{1/3}
	\label{eq:nusseltno_miller},
\end{equation}
and $\xi_\text{sup}$ is the non-dimensional evaporation parameter under the superheated regime given by 
\begin{equation}
       \xi_\text{sup}=\left(\frac{\Prg\ddhat}{ 2\mug}\right)\left[\frac{h_1 (\Thatd-\Thats)}{L(\Thats)}+\frac{f_3h_2 |\Thatg-\Thats|}{L(\Thats)}\right],
\end{equation}
where $\Prg$ is the gas-phase Prandtl number defined as 
\begin{equation}
    \Prg = \frac{\mug(\Tref) \Cpg(\Tref)}{\lambdag(\Tref)}.
    \label{eq:prg}
\end{equation}
For subcooled droplets, the conductive heat flow and the external vaporization are solely responsible for the change in droplet bulk temperature, and thus $\Kheatsub$ can be expressed as
\begin{equation}\label{HeatTransfer2}
    \Kheatsub=\underbrace{\frac{6f_\text{2,sub} \Nud\lambdag(\Tref) (\Thatg - \Thats)}{\rhol \ddhatsq \Cl(\Thatd)}}_\text{conductive heat flow} - \underbrace{\overbrace{\frac{3 \Kvapsub L(\Thats)}{2 \ddhat^2 \Cl(\Thatd)}}^\text{external vaporization}}_\text{evaporative cooling}\hskip0.5cm\text{with}\hskip0.4cm	f_\text{2,sub} = \frac{\xi_\text{sub}}{e^{\xi_\text{sub}}-1}.
\end{equation}
In the subcooled regime, $\Thats$ is considered to be equal to the droplet bulk temperature $\Thatd$. The non-dimensional evaporation parameter in the subcooled regime, $\xi_\text{sub}$, is computed as~\citep{Deshmukh2022}
\begin{equation}
	\xi_\text{sub}  =  \frac{1}{2}\frac{\Prg}{\Scg}\; \mathrm{ln} (1+\BMd) \Shd.
	\label{eq:xi}
\end{equation}
The source term due to heat transfer is thus modeled as 
\begin{equation}
\wheat=\Kheatsup\rhohat\Yhatl.
\end{equation}
\subsection{Nozzle exit conditions}
A blob injection model~\citep{Reitz1987, Reitz1987a} was incorporated by \cite{Deshmukh2022} which injects the droplet with a diameter equivalent to the size of the effective nozzle hole diameter into the gas phase. However, during flash-boiling injection, near-nozzle droplet shattering due to rapid disintegration of the superheated liquid jet results in a significant reduction in droplet diameter at the nozzle exit (as low as $\sim$ 10 $\%$ of the nozzle hole diameter)~\citep{Price2020,Price2016}. In-nozzle phase change due to cavitation may further enhance the near nozzle atomization characteristics of the superheated liquid jet~\citep{Gemci2004}. The use of an initial droplet of size comparable to the nozzle exit diameter at flash-boiling conditions would result in an unrealistic flash-boiling spray with no plume merging or spray collapse. The size of the initial droplet diameter is thus crucial in determining the global characteristics of a flash-boiling spray~\citep{Price2016,Price2015}. In this work, the CAS model has been integrated with a correlation proposed by \cite{Gemci2004} to compute the initial droplet diameter ($\Do$) as
\begin{equation}\label{superheatD}
    \Do=118.40-28.29\left(\DeltaTstar-\text{CN}\right)\hskip0.35cm\text{with}\hskip0.35cm\DeltaTstar=\frac{\Tinj-\Tb\left(\Pg\right)}{\Tb\left(\Pinj\right)-\Tb\left(\Pg\right)}\hskip0.35cm\text{and}\hskip0.35cm\text{CN}=\frac{\Pinj-\Pv\left(\Tinj\right)}{\Pinj-\Pg},
\end{equation}
where $\Pinj$ is the injection pressure, $\DeltaTstar$ the dimensionless superheating degree, and CN the cavitation number. The nozzle exit velocity is computed from the Bernoulli equation considering the losses in the nozzle through the measured discharge coefficient ($C_\text{d}$) as~\citep{Sarre1999}
\begin{equation}
   \Uo=C_\text{d}\sqrt{\frac{2\left(\Pinj-\Pg\right)}{\rhol}}.
\end{equation}

\section{Numerical methodology}\label{Num_Method}
The hyperbolic GEs shown in \myeqs{eq:amb_cont}-\eqref{eq:liq_energy} are non-dimensionalized using the initial droplet diameter ($\Do$) as the length scale, the nozzle exit velocity ($\Uo$) as the velocity scale, and $\tau=\Do/\Uo$ as the time scale. The non-dimensional GEs are then solved numerically in conservative form using the Lax-Friedrichs scheme incorporating Rusanov fluxes~\citep{Rusanov1961} with local wave speeds. The GEs are advanced in time using an explicit Euler scheme. The initialization of the non-dimensional variables in the GEs is performed in the 1D discretized domain with the left boundary treated as the liquid jet and the right boundary as the far-field condition. The initial and boundary conditions used in the CAS model are summarized in \mytab{IC_BC}. For more details on the numerical methods and solution procedure, the reader is referred to \cite{Deshmukh2022}. A Courant-Friedrichs-Lewy (CFL) number of 0.1 is used for all CAS simulations in this work. The CAS model is implemented in an in-house serial FORTRAN90 code framework. 
\renewcommand{\arraystretch}{1.0}
\begin{table}[h]
\begin{center}
\begin{tabular}{c@{\quad}c@{\quad}c@{\quad}c@{\quad}c@{\quad}c}
\toprule
\textbf{Variable} & \textbf{Definition} & \textbf{IC} & \textbf{Left BC} & \textbf{Right BC} & \textbf{Bounds} \\
\midrule
$\Ystarl$ & $\Yhatl$ & 0.0 & Dirichlet 1.0 & Neumann 0.0 & [0.0,1.0] \\
$\Ystarv$ & $\Yhatv$ & 0.0 & Dirichlet 0.0 & Neumann 0.0 & [0.0,1.0] \\
$\Ystara$ & $\Yhata$ & 1.0 & Dirichlet 0.0 & Neumann 0.0 & [0.0,1.0] \\
$\rhostar$ & $\rhohat/\rhol$ & $\rhoa/\rhol$ & Dirichlet 1.0 & Neumann 0.0 & [$\rhog/\rhol$,1.0] \\
$\ustarl$ & $\uhatl/\Uo$ & 0.0 & Dirichlet 1.0 & Neumann 0.0 & [0.0,1.0] \\
$\ustarg$ & $\uhatg/\Uo$ & 0.0 & Dirichlet 0.0 & Neumann 0.0 & [0.0,1.0] \\
$\mean{\ddstar}$ & $\mean{\ddhat}/\Do$ & 0.0 & Dirichlet 0.1-1.0 & Neumann 0.0 & [0.0,1.0] \\
$\Tstard$ & $\Thatd/\Tinj$ & 0.0 & Dirichlet 1.0 & Neumann 0.0 & [0.0,1.0] \\
$\Tstarg$ & $\Thatg/\Tinj$ & $\Tamb/\Tinj$ & Neumann 0.0 & Neumann 0.0 & [$\Tamb/\Tinj$,1.0] \\
$\bstar$ & $b/\Do$ & 0.5 & Dirichlet 0.5 & Neumann 0.0 & [0.5,$\infty$) \\
$\zstar$ & $z/\Do$ & - & 0.0 & 800.0 & [0.0,800.0] \\
$\tstar$ & $t/\tauo$ & 0.0 & - & - & [0.0,$\infty$) \\
\bottomrule
\end{tabular}
\end{center}
\caption{Initial and boundary conditions used in the CAS model.}
\label{IC_BC}
\end{table}
\section{Description of cases for model validation}\label{Case_Des}
The newly developed CAS model results on the evaluation of flash-boiling spray characteristics are validated against the experimental measurements reported by \cite{Aleiferis2013} and \cite{Duronio2021} for two different injector nozzles. The first is a 6-hole asymmetric injector (herein referred to as `Aleiferis injector'), where the Shadowgraphy technique was used to visualize the spray and hence only the liquid phase was measured in their experiments~\cite{Aleiferis2013}. Differently, \cite{Duronio2021} measured both the liquid and vapor phase using Mie scattering and Schlieren techniques, respectively, for the ECN Spray G injector (8-hole asymmetric)~\citep{ECN2020}. The geometric details of both injectors are listed in \mytab{InjDet}. A detailed description of the cases selected for model validation is summarized in \mytab{Test_Cases}. The tuned values of the breakup model constants for different injector nozzles are shown in \mytab{BrkConst}. The injection rate profiles for all the investigated cases are generated using the virtual injection rate generator provided by \citep{CMT2023}.
\renewcommand{\arraystretch}{1.0}
\begin{table}[!t]
  \begin{center}
\def~{\hphantom{0}}
  \begin{tabular}{c@{\quad}c@{\quad}c@{\quad}c@{\quad}c@{\quad}c@{\quad}c}
  \toprule
  &&&\multicolumn{2}{c}{\textbf {Injectors}}\\ 
  \cline{4-5}
       \multirow{-2}{*}{\textbf{Parameter}}                      &   \multirow{-2}{*}{\textbf{Symbol}}  &    \multirow{-2}{*}{\textbf{Unit}}  & \textbf{Aleiferis}  & \textbf{Spray G}\\
  \midrule 
  \# no of holes                             & - & - & 6 & 8         \\
Orifice diameter                             & $\Dnoz$ & \si{\micro\metre} & 200 & 165         \\
Length to diameter ratio                  & $\Lnoz/\Dnoz$ & -  & 1.0$-$1.1 & 1.4       \\ 
Outer spray angle  & $\theta_1$ &    \si{\degree}     & 60 & 80 \\ 
Discharge coefficient                  & $C_\mathrm{d}$  & - &  0.6 & 0.64    \\
  \bottomrule
  \end{tabular}
  \caption{Geometric details of the different injector nozzles.}
  \label{InjDet}
  \end{center}
\end{table}
\renewcommand{\arraystretch}{1.0}
\begin{table}[!h]
\begin{center}
\begin{tabular}{c@{\quad}c@{\quad}c@{\quad}c@{\quad}c@{\quad}c@{\quad}c@{\quad}c@{\quad}c@{\quad}c@{\quad}c@{\quad}c@{\quad}c}
\toprule
\textbf{No} $\downarrow$ &\textbf{Case} & \textbf{Injector} & \textbf{Fuel} & $\Pinj$ & $\Tinj$ & $\Pg$ & $\Tg$  & $\DeltaT$ \\ 
\midrule 
Unit $\rightarrow$& - & - & - & \si{\bar} & \si{\kelvin} & \si{\bar} & \si{\kelvin}  & \si{\kelvin} \\
\midrule 
1&`PEN54' & Aleiferis & \pen & 150 & 363.15 & 1.0 & 298.15  & 54\\
2&`PEN84' & Aleiferis & \pen & 150 & 393.15 & 1.0 & 298.15 & 84\\
3&`PEN103' & Aleiferis & \pen & 150 & 393.15 & 0.5 & 298.15 & 103\\
4&`ETH29' & Aleiferis & ethanol & 150 & 363.15 & 0.5 & 298.15  & 29  \\
5&`ETH42' & Aleiferis & ethanol & 150 & 393.15 & 1.0 & 298.15 & 42  \\
7&`OCT44' & Aleiferis & \isooct & 150 & 393.15 & 0.5 & 298.15  & 44 \\
6&`G200' & Spray G & \isooct & 200 & 363.15 & 0.2 & 333.15 & 39 \\
8&`G150' & Spray G & \isooct & 150 & 363.15 & 0.2 & 333.15  & 39 \\
9&`G100' & Spray G & \isooct & 100 & 363.15 & 0.2 & 333.15  & 39 \\
\bottomrule
\end{tabular}
\end{center}
\caption{Test cases chosen for model validation. The values of $\DeltaT$ are given to their nearest degree. The breakup model constants are tuned for injector nozzles for cases no. 1 and 6.}
\label{Test_Cases}
\end{table}
\renewcommand{\arraystretch}{1.0}
\begin{table}[!h]
\begin{center}
\begin{tabular}{c@{\quad}c@{\quad}c@{\quad}c}
\toprule 
\textbf{Model} & \textbf{Constant}  & \multicolumn{2}{r}{\textbf{Present work}} \\ 
\midrule 
Kelvin-Helmholtz & $B_0$  &  \multicolumn{2}{r}{0.61} \\
& {$B_1$}  &  & 5.0 \\
\midrule
\multirow[t]{4}{*}{Rayleigh-Taylor} & \multirow[t]{4}{*}{$C_3$} &  & 0.45 \\
\midrule
\multirow[t]{4}{*}{Thermal breakup} & \multirow[t]{4}{*}{$A_0$}  & Aleiferis: & 1.0 \\
& &  Spray G:& 0.5\\
& \multirow[t]{4}{*}{$A_1$}  & Aleiferis: & 0.85 \\
& &  Spray G:& 0.75 \\
\bottomrule
\end{tabular}
\end{center}
\caption{Tuned breakup model constants for Aleiferis and Spray G injectors.}
\label{BrkConst}
\end{table}
\section{Results and discussion}\label{results}
In this section, first, the performance of the CAS model without the model extensions is evaluated with the experimental measurements for different fuels and operating conditions, as listed in \mytab{Test_Cases}. Then, the improvements in the CAS model predictions with the proposed model extensions are discussed. Since the liquid and vapor penetration lengths are considered as the most important performance metric in determining the spray characteristics~\citep{Pickett2011, Siebers1999}, the simulation results are compared in terms of the penetration lengths, which are calculated based on the ECN guidelines~\citep{ECN2020}. 
\subsection{Results without model extensions}\label{CAS_results}
\begin{figure}[!t]
\centering
\includegraphics[width=410pt]{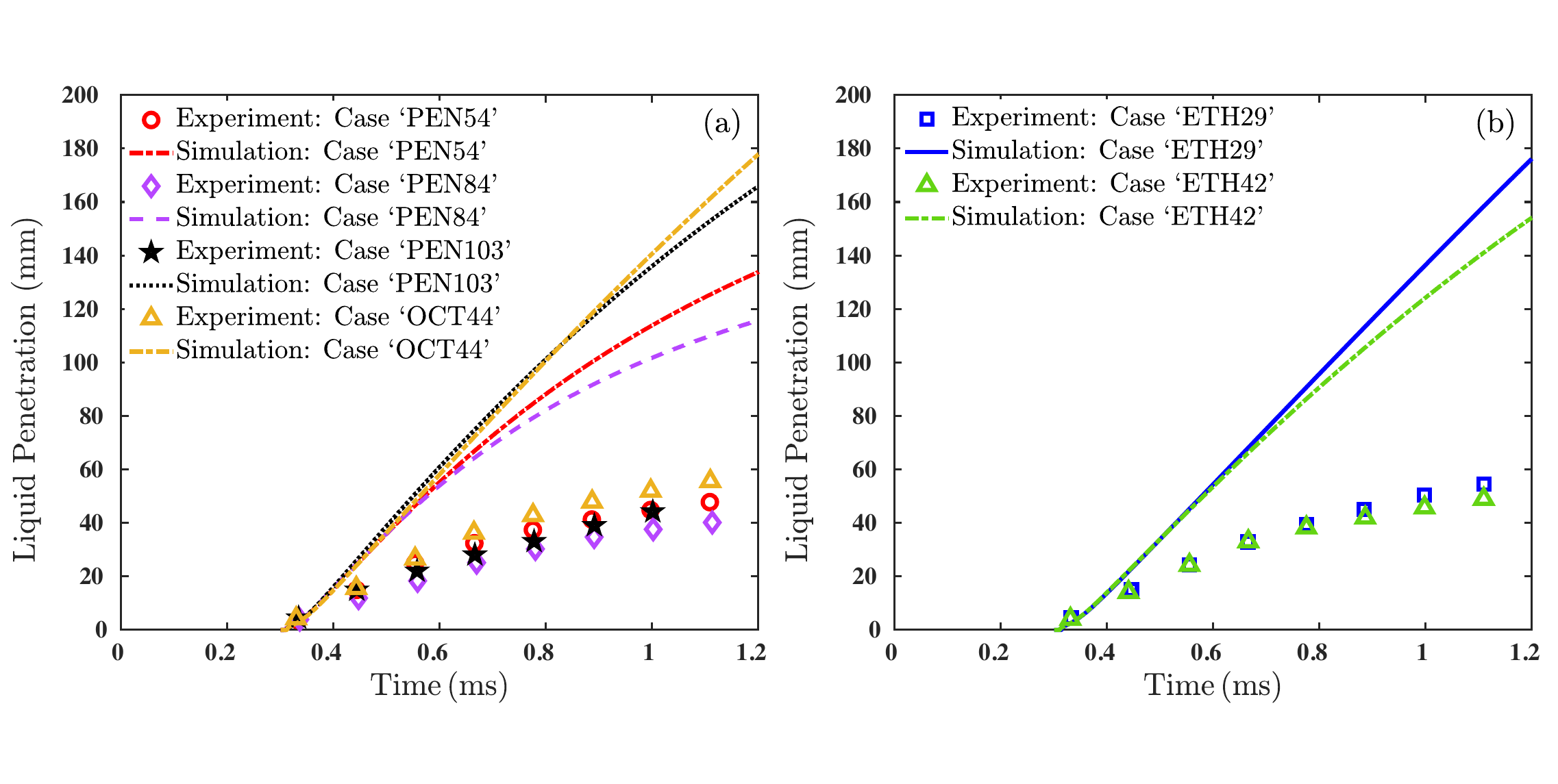}   
\includegraphics[width=410pt]{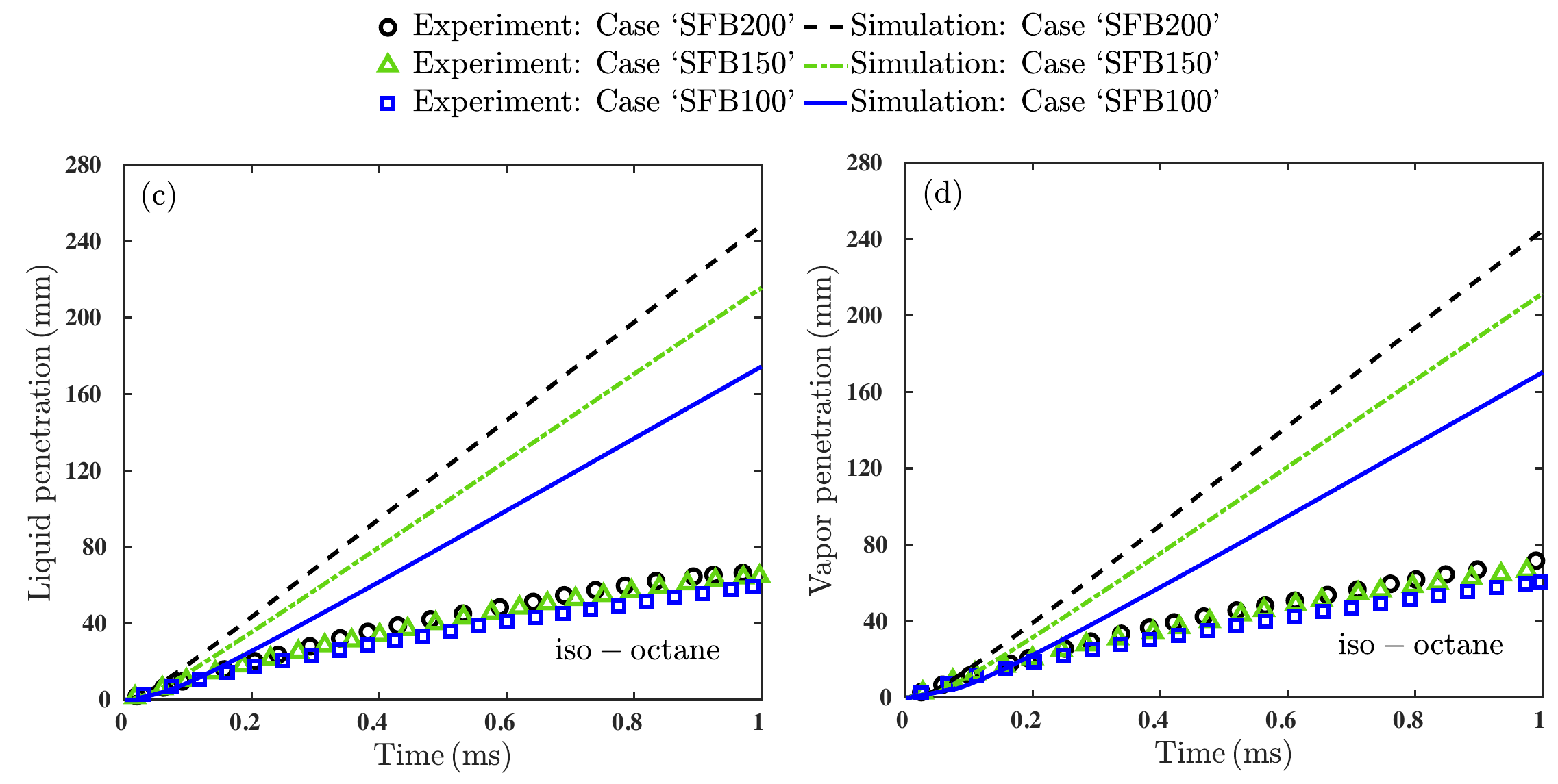}   
\vspace{0.1 in}
\caption{Comparison of the spray characteristics predicted by the CAS model without the model extensions and the experiments. Subfigures (a) \& (b) show the comparison of liquid penetration lengths for the Aleiferis injector~\citep{Aleiferis2013} under different operating conditions for three different fuels, whereas subfigures (c) \& (d) illustrate the comparison of liquid and vapor penetration lengths of \isooct~fuel, respectively, at different injection pressures for the ECN Spray G injector~\citep{Duronio2021}.}
\label{PenCAS}
\end{figure}

The spray penetration lengths predicted by the CAS model without the model extensions described in \mysec{CAS_model} are compared with the experimental measurements in \myfig{PenCAS} for different fuels and operating conditions. The detailed simulation parameters of the investigated cases are listed in \mytab{Test_Cases}. It is observed that the physical submodels used by \cite{Deshmukh2022} are unable to reproduce the enhanced evaporation and atomization behavior of flash-boiling sprays, thus resulting in significant over-prediction in the spray penetrations for both injectors. The model even fails to capture the trends in spray characteristics for \pen~and ethanol fuels. \myfig{PenCAS}a depicts that for \pen~fuel, the predicted penetration length increases with increasing superheating degree from $\DeltaT=84$ K (case `PEN84') to 103 K (case `PEN103'), whereas a decreasing trend was observed in the experiments. A similar increasing trend is also predicted for ethanol with increasing $\DeltaT$ from 29 K (case `ETH29') to 42 K (case `ETH42'), which is in contradiction with the experimental measurements. These emphasize the importance of the model extension for accurately predicting the spray characteristics under flash-boiling conditions.\par 
\subsection{Results with model extensions}
This section discusses the gradual enhancements made in the previous CAS model performance, highlighting the inclusion or substitution of the new sub-models, as described in \mysec{CAS_model}.
\subsubsection{Spray cone angle and initial droplet size models}
\begin{figure}[!b]
\centering
\includegraphics[width=410pt]{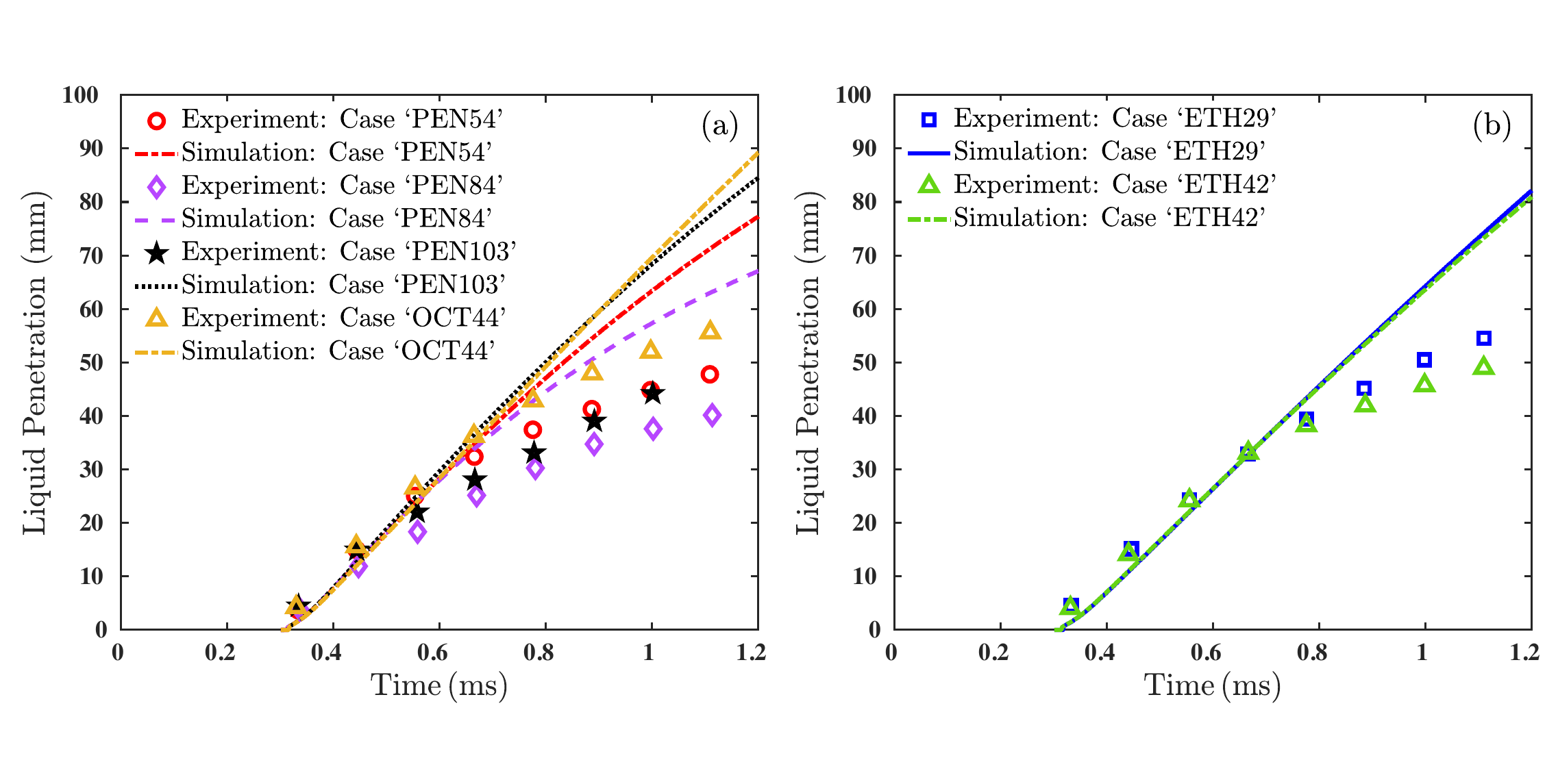}   
\includegraphics[width=410pt]{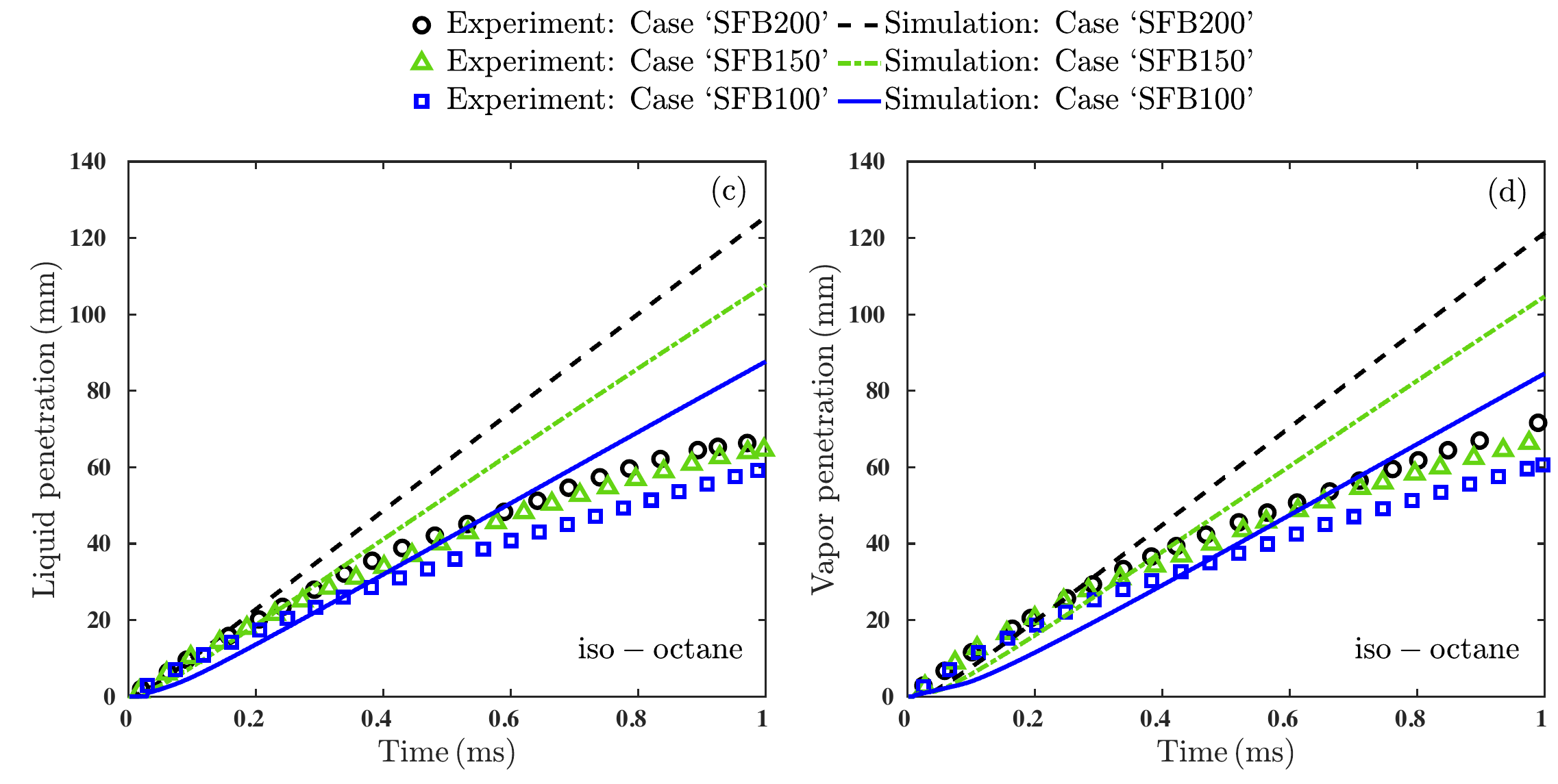}   
\vspace{0.1 in}
\caption{Spray characteristics predicted by the CAS model with updated spray cone angle (\myeq{Cone}) and initial droplet size (\myeq{superheatD}) models. Subfigures (a) \& (b) describe the comparison of liquid penetration lengths for the Aleiferis injector~\citep{Aleiferis2013} under different operating conditions for three different fuels, whereas subfigures (c) \& (d) illustrate the comparison of liquid and vapor penetration lengths of \isooct~fuel, respectively, at different injection pressures for the ECN Spray G injector~\citep{Duronio2021}.}
\label{CAS2}
\end{figure}
Flash-boiling sprays are characterized by a widening of the spray plume due to the increased radial expansion resulting from the bubble growth and micro-explosion~\citep{Price2018}. Thus, accurate predictions of the spray cone angle as well as the initial droplet diameter play a crucial role in determining the global spray characteristics. \myfig{CAS2} shows the CAS model performance with the upgraded models for spray cone angle (\myeq{Cone}) and initial droplet size (\myeq{superheatD}) in comparison with the previous version during flash-boiling conditions. Here, the droplet heat transfer and vaporization processes are modeled via the standard Miller and Bellan model. For the droplet breakup, only KH--RT breakup model is incorporated. The gas phase temperature is obtained via the approach considered by \cite{Deshmukh2022}. It is observed from \myfig{CAS2} that the predictive capabilities of the CAS model are significantly improved with the updated cone angle and initial droplet diameter models compared to the previous version.  However, the penetration lengths are quantitatively still overpredicted compared to the experimental measurements.  
\subsubsection{Superheated droplet vaporization, heat transfer, and breakup models}
\begin{figure}[h]
\centering
\includegraphics[width=410pt]{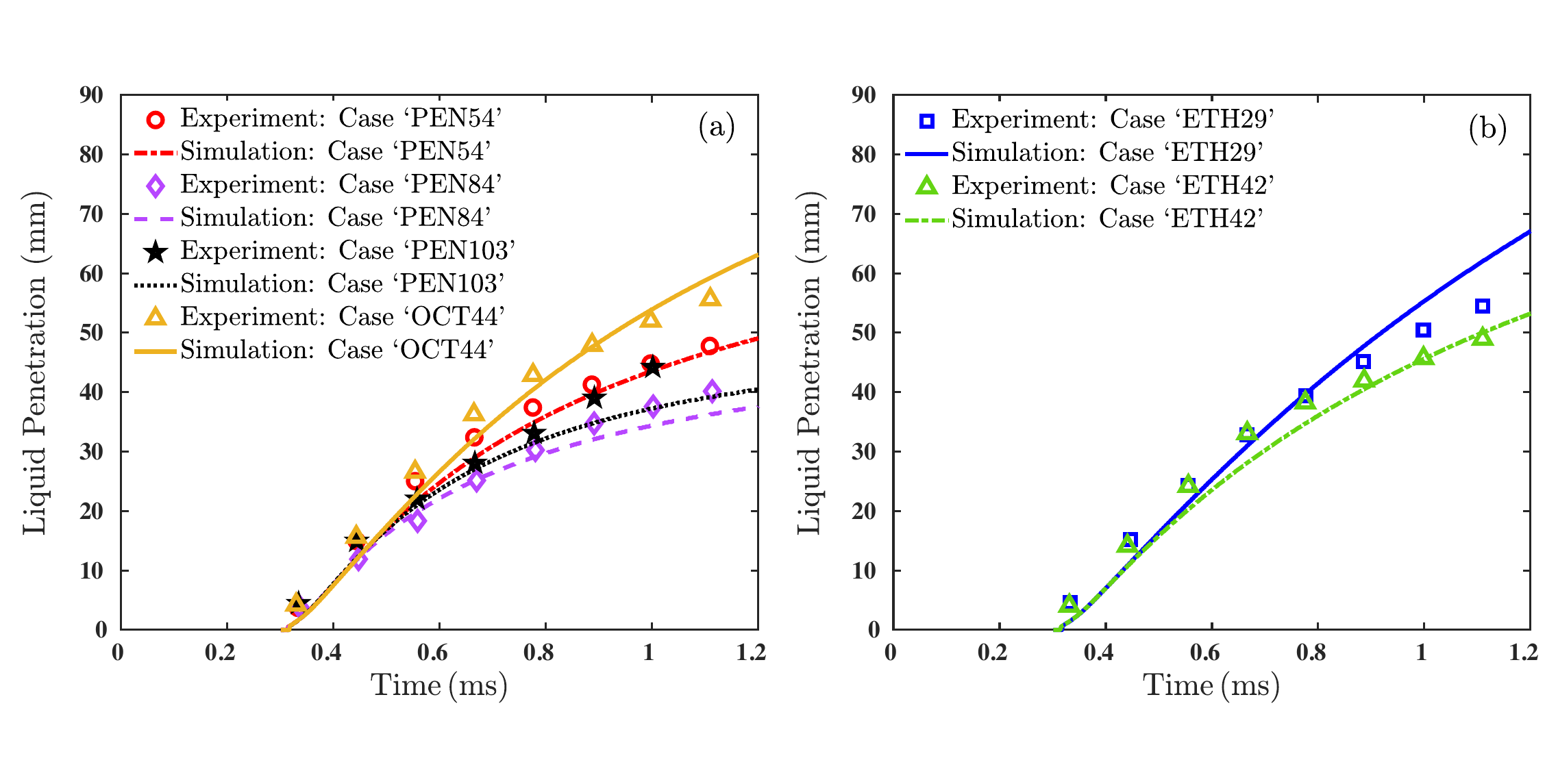}   
\vspace{-0.2 in}
\caption{Spray characteristics predicted by the upgraded CAS model. Subfigures (a) \& (b) describe the comparison of liquid penetration lengths for the Aleiferis injector~\citep{Aleiferis2013} for three different fuels under different operating conditions, as listed in \mytab{Test_Cases}.}
\label{PenEth}
\end{figure}

This section concludes the extension of the CAS model with the upgrade of the standard evaporation, heat transfer, and breakup models to that of the superheated vaporization, heat transfer, and hybrid aerodynamic-thermal breakup models, as described in \mysec{CAS_model}. Here, \myeq{eq:gas_energy} is incorporated for the calculation of the gas phase temperature. The improvements in the spray characteristics obtained from the fully updated CAS model are shown in \myfig{PenEth} for different fuels under varying superheating degrees for the Aleiferis injector. 
\myfig{PenEth}a shows that the predicted spray penetration decreases with increasing $\Tinj$, from 363.15 K (case `PEN54') to 393.15 K (case `PEN84'), at constant $\Pg=1.0$ bar for \pen~fuel, as observed in the experiments. The increase in $\Tinj$ leads to an increase in liquid superheating degree ($\DeltaT$) for the case `PEN84', thus enhancing the evaporation rate and thermally induced breakup~\citep{Yang2013}, which in turn produces the smaller droplets. These droplets then substantially decelerate due to the aerodynamic drag forces of the ambient gas, resulting in shorter spray penetrations. A similar trend is observed for flash-boiling ethanol (case `ETH29' and `ETH42'), as illustrated in \myfig{PenEth}b.\par
With the system pressure $\Pg$, decreasing from 1.0 bar (case `PEN84') to 0.5 bar (case `PEN103'), at $\Tinj=393.15$ K for \pen~fuel, \myfig{PenEth}a shows that the predicted penetration length increases, as also reported in the experimental findings by \cite{Aleiferis2013}. \mytab{Test_Cases} shows that decreasing the system pressure for these \pen~test cases leads to an increase in $\DeltaT$ from 84 K to 103 K. The reason behind this phenomenon is the lower saturation temperature of the liquid under reduced pressure conditions. As described above, the liquid at higher $\DeltaT$ is associated with increased evaporation and enhanced atomization effects, thus expected to result in smaller spray droplets and eventually shorter spray penetrations compared to the case with lower $\DeltaT$. However, the resulting opposite trend in penetrations for these \pen~test cases can be attributed to the lower system pressures. By lowering the pressures, the reduced aerodynamic drag forces are likely to outweigh the enhanced evaporation and atomization effects at high $\DeltaT$, leading to an increase in penetration lengths. The predicted penetration length for \isooct~fuel at $\DeltaT=44$ K (case `OCT44') also agrees well with the experiment, as shown in \myfig{PenEth}a. Overall, the present CAS model is able to predict the trends in macroscopic spray characteristics similar to the experiments for different fuel properties and operating conditions for a given injector. \par
\begin{figure}[!b]
\centering
\includegraphics[width=400pt]{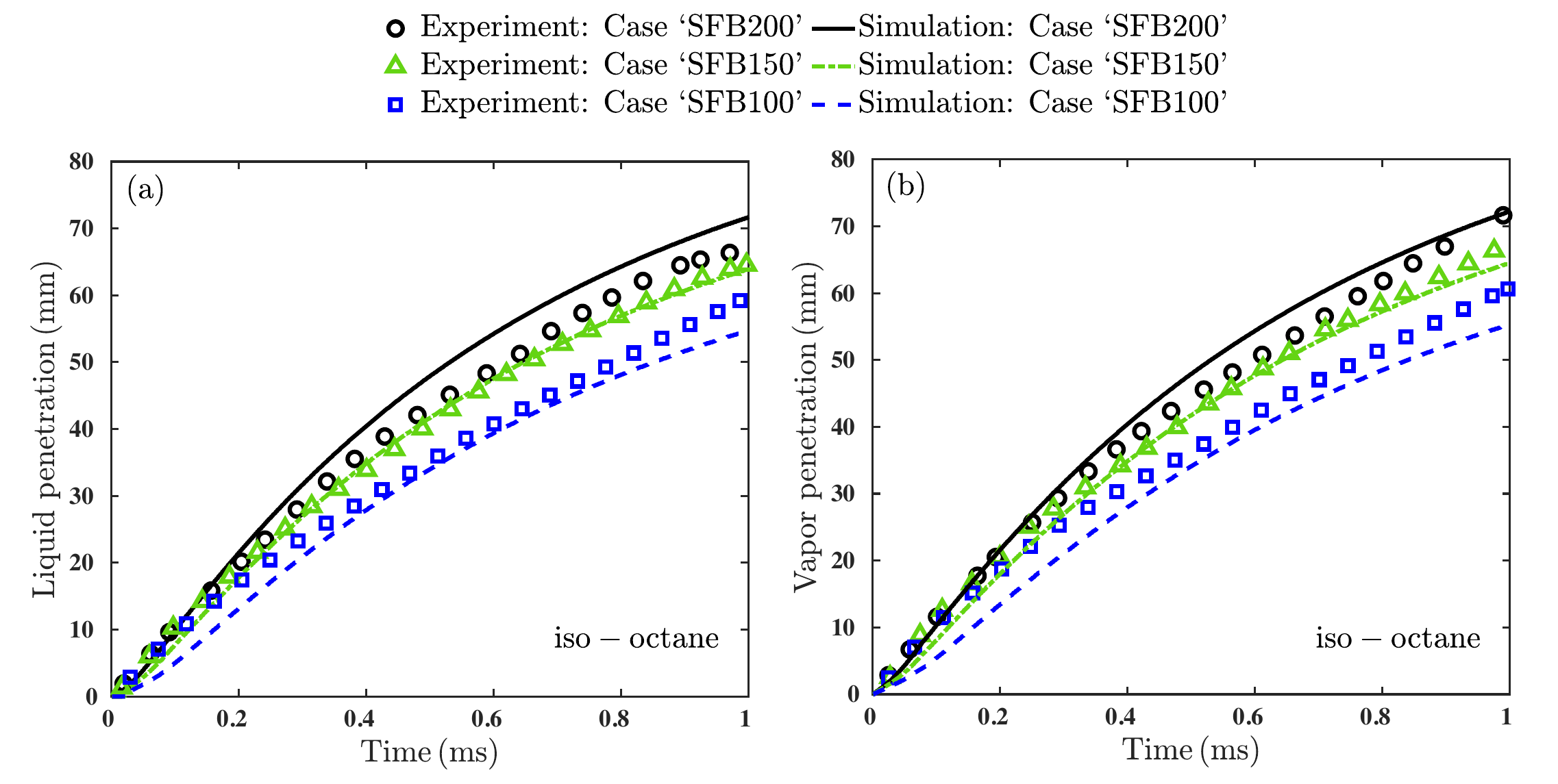}  
\caption{Spray characteristics predicted by the upgraded CAS model. Subfigures (a) \& (b) illustrate the comparison of liquid and vapor penetration lengths of \isooct~fuel, respectively, for the ECN Spray G injector~\citep{Duronio2021} at different injection pressures, as listed in \mytab{Test_Cases}.}
\label{PenEth_CAS2}
\end{figure}

\myfig{PenEth_CAS2} depicts the improved spray characteristics of the ECN Spray G injector for varying injection pressures at a constant $\Rp$ of 0.26. The decreasing injection pressure leads to a reduction in the mass flow rate and subsequently, the residence time within the nozzle hole will be increased. Due to the longer residence time, the vapor bubbles start nucleating inside the injector nozzle leading to the formation of a well-atomized spray with a shorter penetration length. It is observed that the extended CAS model is able to predict the decreasing trend of spray penetrations with decreasing injection pressures, as also observed in the experiments. \par
\begin{figure}[!b]
\centering
\includegraphics[width=410pt]{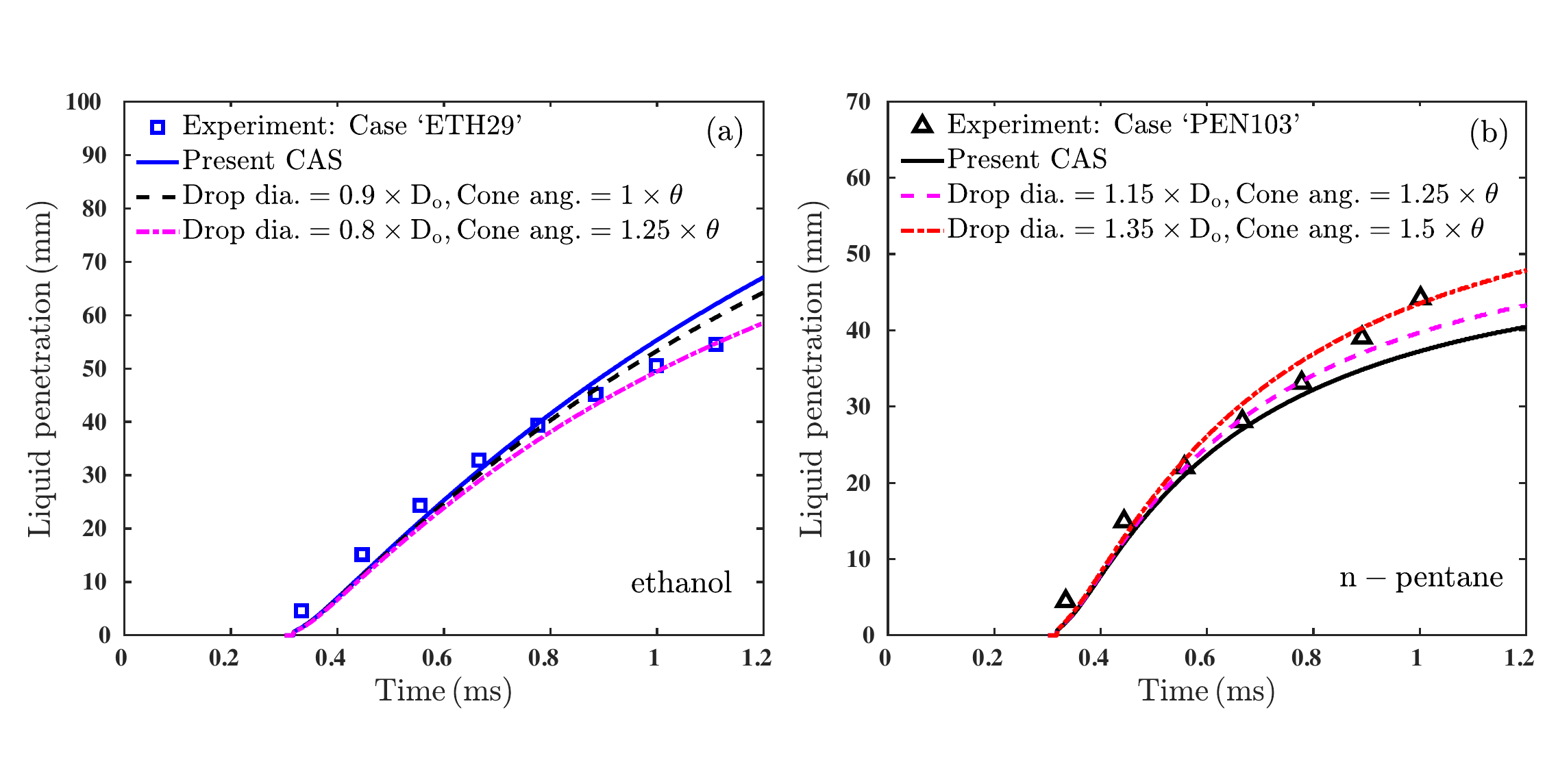} 
\vspace{-0.15 in}
\caption{Influence of the spray cone angle and initial droplet diameter on the liquid penetration for (a) un-collapsed and (b) collapsed sprays of ethanol and \pen~fuels, respectively.}
\label{Inf}
\end{figure}

However, quantitatively, for some test cases, the present CAS model still over- and/or under-predicts the penetration lengths for both injectors. Although this is expected due to the averaged approach of the CAS model, the use of more accurate models for calculating the spray cone angle and initial droplet diameter is expected to improve the prediction towards the experiments vastly. This is because the increased radial expansion resulting from the bubble growth and micro-explosion would widen the spray plume as well as reduce initial droplet sizes under flash-boiling conditions~\citep{Price2018}. For a multi-hole injector, the wider spray plumes emerging from different injector holes could easily merge with each other depending on the inter-spacing and directions of the holes, thus resulting in even higher total plume widening compared to a single-hole injector. The present CAS model would not be able to capture the shattering of the child droplets due to micro-explosion and subsequent widening of the spray plume and reduced droplet sizes because of the continuous breakup approach considered. Thus, the spray cone angle and the initial droplet size models used in this study play a crucial role to mimic the above-mentioned phenomena. The influence of $\theta$ and $\Do$ on the penetration length is illustrated in \myfig{Inf}a and \myfig{Inf}b for un-collapsed (case `ETH-29') and collapsed (case `PEN103') sprays, respectively. It can be seen that the reduced initial droplet size and higher spray cone angle improve the quantitative prediction of liquid penetration for the un-collapsed spray, whereas the increased droplet diameter associated with higher spray cone angle provides a more reasonable prediction of the liquid penetration length for the fully collapsed spray. \par
\subsection{Computational cost}
In order to investigate the cost reduction factor associated with using the CAS model in comparison to the 3D simulations, an LES of flash-boiling spray is performed for case `PEN54'. A brief description of the numerical solver used for LES and the simulation results are included in \ref{LES}. Both the 1D and 3D simulations were run on machines that use the Intel Broadwell processor architecture. The 3D LES was run on 240 cores for $\approx$ 16.7 h leading to a total of 4008 CPUh. The 1D CAS was run on the same machine on 1 core for $\approx$ 205 s resulting in a total of 0.057 CPUh. Thus, the CAS model is faster by up to 4 orders of magnitude compared to the 3D LES while providing reasonable predictions in flash-boiling spray characteristics such as liquid and vapor penetration lengths for different operating conditions and fuels. 
\section{Conclusions}\label{concl}
A reduced-order cross-sectionally averaged flash-boiling spray model was proposed in this work. The main conclusions of the present study can be summarized as follows: 
\begin{itemize}
\item The previously developed CAS model was first applied to predict the spray characteristics such as penetration lengths for different fuels under flash-boiling conditions. It was found that the CAS model fails to reproduce the trend in the flash-boiling spray penetration lengths. 
\item An extension of the CAS model was then proposed to improve its predictive capabilities for the simulation of flash-boiling sprays. The important physical subprocesses in flash-boiling sprays such as internal vaporization, external vaporization, and thermally driven breakup were incorporated into the newly developed CAS model. The initial droplet diameter for flash-boiling sprays, which is expected to be considerably smaller than the nozzle exit diameter, was estimated using an experimental correlation available in the literature. Additionally, an appropriate empirical formulation was employed to model the wider spray plume angle observed in flash-boiling conditions 
\item The upgraded CAS model performance was compared with the experimental measurements of two different injector nozzles for varying injection pressures and superheating degrees. It was found that the trends in liquid and vapor penetration lengths predicted by the updated CAS model agree well with the experiments. 
\item Though in some cases, the penetration lengths were found to be under-and/or over-predicted due to the averaged approach of the CAS model, it was shown that the prediction could be improved with the more accurate modeling of the spray cone angle and initial droplet diameter. 
\item A 3D LES of a flash-boiling spray was also performed in order to assess the computational efficiency of the CAS model. The 1D CAS model was shown to be faster by up to four orders of magnitude in comparison to the 3D LES, thus making it really useful in many practical applications related to flash-boiling including but not limited to the design of experiments, rapid fuel-screening, and creating digital twins.   
\end{itemize}

\section*{Acknowledgements}
This work was performed as part of the Cluster of Excellence “The Fuel Science Center”, which is funded by the Deutsche Forschungsgemeinschaft (DFG, German Research Foundation) under Germany’s Excellence Strategy – Exzellenzcluster 2186
“The Fuel Science Center” ID: 390919832.
\appendix
\section{Derivation of the superheated vaporization coefficients}\label{App_evap}
subsection{Internal Vaporization}
The vaporization of the superheated liquid causes the vapor bubbles to grow in size and subsequently leads to the expansion of the liquid droplet. For a single bubble, from the mass continuity 
\begin{equation}\label{BubGrth}
    \rhol\frac{\text dR_\text d}{\text dt}=\rho_\text v\frac{\text dR_\text b}{\text dt}.
\end{equation}
Rearranging \ref{BubGrth} yields
\begin{equation}\label{SingBub}
        \frac{\text dd^2_\text d}{\text dt}=
    4d_\text d\frac{\rho_\text v}{\rhol}\frac{\text dR_\text b}{\text dt}.
\end{equation}
Summing over the total number of vapor bubbles, \ref{SingBub} becomes
\begin{equation}\label{MultiBub}
        \frac{\text dd^2_\text d}{\text dt}=
    \Nbub\times4d_\text d\frac{\rho_\text v}{\rhol}\frac{\text dR_\text.pdf b}{\text dt}=\Kvapswl
\end{equation}
\subsection{External Vaporization}
The vapor mass flux from the droplet outer surface due to the heat transfer from the droplet inner core is modeled as~\citep{Adachi1997}
\begin{equation}\label{IntMass}
        \rho_\text sv_\text s=\frac{h_\text{f}\left(\Td-\Tb\right)}{L\left(\Tb\right)},
\end{equation}
where $v_\mathrm{s}$ denotes the vapor flow rate from the droplet outer surface. From the mass continuity at the droplet surface
\begin{equation}\label{MasCon1}
        \frac{\rhol}{2}\frac{\mathrm d\dd}{\mathrm dt}=\rho_\text sv_\text s.
\end{equation}
Equating \myeq{MasCon1} with the \myeq{IntMass} yields
\begin{equation}\label{MasCon2}
        \frac{\rhol}{2}\frac{\mathrm d\dd}{\mathrm dt}=\frac{h_\text{f}\left(\Td-\Tb\right)}{L\left(\Tb\right)}.
\end{equation}
Rearranging \myeq{MasCon2}, the rate of change of $\ddsq$ can be expressed as
\begin{equation}
        \frac{\mathrm{d}\ddsq}{\mathrm{d}t}=\frac{4\dd h_\text{f}\left(\Td-\Tb\right)}{\rhol L\left(\Tb\right)}=\Kvapint.
\end{equation}
Similarly, the vapor mass flux due to the temperature gradient between the droplet surface and the external ambient is expressed as 
\begin{equation}\label{ExtMass}
        \rho_sv_s=\frac{h_\text{ex}f_3\left|\Tg-\Ts\right|}{L\left(\Ts\right)}.
\end{equation}
Using mass continuity at the droplet surface (\myeq{MasCon1}) and rearranging \myeq{ExtMass} yields
\begin{equation}
    \frac{\mathrm{d}\ddsq}{\mathrm{d}t}=\frac{4 \dd f_3h_\mathrm{ex}\left|\Tg-\Ts\right|}{\rhol L\left(\Ts\right)}=\Kvapext.
\end{equation}
\section{Derivation of the heat transfer coefficient}\label{App_heat}
The energy balance of the droplet-bubble system in the superheated regime can be written as
\begin{equation}\label{EngBal}
   m_\text dC_\text l\left(\Td\right)\frac{\mathrm{d}\Td}{\mathrm{d}t}=4\pi r^2_\text dQ_\text d-4\pi r^2_\text d\rho_\text s v_\text sL\left(\Ts\right)-4\pi R^2_\text b\rhov\Nbub\dot R_\text bL\left(\Td\right).
\end{equation}
Substituting 
\begin{equation}
Q_{\mathrm{d}}=\frac{\lambda_{\mathrm{g}}\left(T_{\mathrm{ref}}\right)\left(\Tg-\Ts\right)}{{\dd}} f_\text{2,sup}\Nud
\end{equation}
and 
\begin{equation}
\rho_s v_s=\frac{h_\text{f}\left(\Td-\Tb\right)}{L\left(\Tb\right)}+\frac{f_3h_\text{ex}|\Tg-\Ts|}{L\left(\Tb\right)}
\end{equation}
in \myeq{EngBal} and rearranging yields
\begin{equation}
   \frac{\text d\Td}{\text dt}=\frac{6f_\text{2,sup} \Nud\lambdag(\Tref) (\Tg - \Ts)}{\rhol \ddsq \Cpl(\Td)}-\frac{3\Kvapsup L\left(\Ts\right)}{2d^2_\text dC_\text l\left(\Td\right)}-\frac{3\Kvapswl L\left(\Td\right)}{2d^2_\text dC_\text l\left(\Td\right)}=\Kheatsup.
\end{equation}
The energy balance of the droplet-bubble system in the subcooled regime can be written as
\begin{equation}\label{EngBal2}
   m_\text dC_\text l\left(\Td\right)\frac{\mathrm{d}\Td}{\mathrm{d}t}=4\pi r^2_\text dQ_\text d-4\pi r^2_\text d\rho_\text s v_\text sL\left(\Ts\right).
\end{equation}
Substituting 
\begin{equation}
Q_{\mathrm{d}}=\frac{\lambda_{\mathrm{g}}\left(T_{\mathrm{ref}}\right)\left(\Tg-\Ts\right)}{{\dd}} f_\text{2,sub}\Nud\hskip0.5cm\text{and}\hskip0.5cm  \rho_\text s v_\text s=2\rho_\text s\Gamma_\text{v,g}\frac{ln(1+B_\text{M,d})}{d_\text d}
\end{equation}
in \myeq{EngBal2} and rearranging yields
\begin{equation}
   \frac{\mathrm{d}\Td}{\mathrm{d}t}=\frac{6f_\text{2,sub} \Nud\lambdag(\Tref) (\Tg - \Ts)}{\rhol \ddsq \Cpl(\Td)}-\frac{3\Kvapsup L\left(\Ts\right)}{2d^2_\text dC_\text l\left(\Td\right)}=\Kheatsub.
\end{equation}
\section{LES of flash-boiling spray}\label{LES}
An LES of a flash-boiling spray is performed using a two-way coupled 3D Lagrangian-Eulerian framework. The in-house code CIAO is used to solve the compressible Navier-Stokes equations (NSEs). CIAO is a structured, high-order, finite-difference code, which solves the NSEs using central difference schemes. For time-marching, CIAO uses a low-storage five-stage, explicit Runge-Kutta (RK) scheme. The subgrid stresses are modeled using a dynamic Smagorinski subfilter model considering Lagrangian averaging~\cite{Germano1991}. For more details about the flow solver in CIAO, the reader is referred to \cite{Mittal2014}. For the computation of the dispersed liquid phase, the Lagrangian equations governing the single droplet position, velocity, mass, and temperature are solved~\cite{Miller1999}. Both the internal and external vaporization processes of the superheated droplets are considered. The newly developed semi-analytical solution for bubble growth rate by \cite{Saha2023} is incorporated to obtain the bubble growth dynamics in the superheated droplets. Due to the smaller size of the droplets in flash-boiling sprays, the conductive thermal resistance in the superheated liquid droplets is neglected and the droplet bulk temperature is modeled using an infinite conductivity model \citep{Miller1999}. The vapor contained in the bubbles is assumed to be in equilibrium with the surrounding superheated liquid medium. A hybrid breakup model consisting of flash-boiling induced breakup~\citep{Senda1994} and aerodynamic breakup~\citep{Patterson1998} is incorporated to simulate the breakup process under superheated conditions.\par
\begin{figure}[!h]
\centering
\includegraphics[width=410pt]{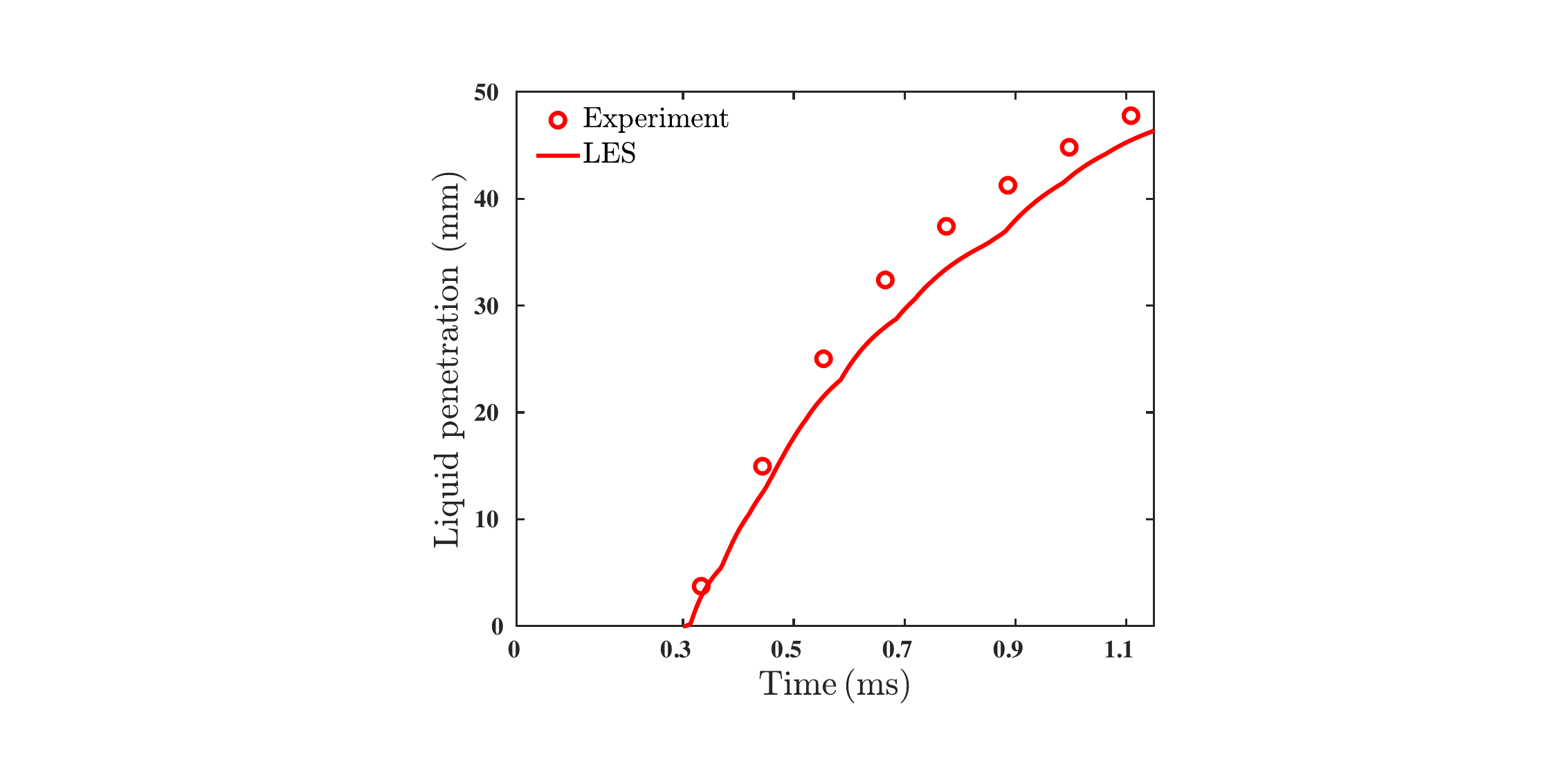}  
\vspace{-0.3 in}
\caption{Comparison of the liquid penetration length of \pen~fuel at $\Pinj=150$ bar, $\Pg=1.0$ bar, and $\Tinj=363.15$ K. Experimental measurements are taken from \cite{Aleiferis2013}.}
\label{LES_Sim}
\end{figure}

The LES is performed on a structured grid with a cell size of 180 $\mu$m resulting in a total cell count of 8.45 million. The simulation results are compared with the experiments in terms of the penetration length and the steady-state Sauter-mean diameter (SMD). \myfig{LES_Sim} shows the comparison of the liquid penetration lengths obtained from the LES and the experimental measurements of \cite{Aleiferis2013} for \pen~fuel at $\Pinj=150$ bar, $\Pg=1.0$ bar, and $\Tinj=363.15$ K.
Considering the complexity of flash-boiling modeling, the spray penetration length obtained from the LES shows reasonable agreement with the experiment. A steady-state SMD value of 7.6 $\mu$m is obtained from the LES, which is also found to be within a few microns of the experimentally measured value of 7.9 $\mu$m. The SMD was measured using a Laser diffraction technique at 30 mm along the injector central axis downstream of the nozzle exit in the experiment. 

\section{Availability of the code }\label{Code}
The FORTRAN90 code framework is made open-source and can be found here: \hyperlink{blue}{https://git.rwth-aachen.de/avijitsaha021/cross-sectionally-averaged-spray-model/}

\appendix

\bibliography{CAS_IJMF/texsrc/references}

\begin{thebibliography}{84}
\expandafter\ifx\csname natexlab\endcsname\relax\def\natexlab#1{#1}\fi
\providecommand{\url}[1]{\texttt{#1}}
\providecommand{\href}[2]{#2}
\providecommand{\path}[1]{#1}
\providecommand{\DOIprefix}{doi:}
\providecommand{\ArXivprefix}{arXiv:}
\providecommand{\URLprefix}{URL: }
\providecommand{\Pubmedprefix}{pmid:}
\providecommand{\doi}[1]{\href{http://dx.doi.org/#1}{\path{#1}}}
\providecommand{\Pubmed}[1]{\href{pmid:#1}{\path{#1}}}
\providecommand{\bibinfo}[2]{#2}
\ifx\xfnm\relax \def\xfnm[#1]{\unskip,\space#1}\fi
\bibitem[{Adachi et~al.(1997)Adachi, McDonell, Tanaka, Senda and
  Fujimoto}]{Adachi1997}
\bibinfo{author}{Adachi, M.}, \bibinfo{author}{McDonell, V.G.},
  \bibinfo{author}{Tanaka, D.}, \bibinfo{author}{Senda, J.},
  \bibinfo{author}{Fujimoto, H.}, \bibinfo{year}{1997}.
\newblock \bibinfo{title}{Characterization of fuel vapor concentration inside a
  flash boiling spray}, in: \bibinfo{booktitle}{International Congress \&
  Exposition}, \bibinfo{publisher}{SAE International}.
\newblock \DOIprefix\doi{https://doi.org/10.4271/970871}.
\bibitem[{Aleiferis et~al.(2010a)Aleiferis, Serras-Pereira, Augoye, Davies,
  Cracknell and Richardson}]{Aleiferis2010a}
\bibinfo{author}{Aleiferis, P.}, \bibinfo{author}{Serras-Pereira, J.},
  \bibinfo{author}{Augoye, A.}, \bibinfo{author}{Davies, T.J.},
  \bibinfo{author}{Cracknell, R.F.}, \bibinfo{author}{Richardson, D.},
  \bibinfo{year}{2010}a.
\newblock \bibinfo{title}{Effect of fuel temperature on in-nozzle cavitation
  and spray formation of liquid hydrocarbons and alcohols from a real-size
  optical injector for direct-injection spark-ignition engines}.
\newblock \bibinfo{journal}{International Journal of Heat and Mass Transfer}
  \bibinfo{volume}{53}, \bibinfo{pages}{4588--4606}.
\newblock \DOIprefix\doi{10.1016/j.ijheatmasstransfer.2010.06.033}.
\bibitem[{Aleiferis et~al.(2010b)Aleiferis, Serras-Pereira, {van Romunde},
  Caine and Wirth}]{Aleiferis2010b}
\bibinfo{author}{Aleiferis, P.}, \bibinfo{author}{Serras-Pereira, J.},
  \bibinfo{author}{{van Romunde}, Z.}, \bibinfo{author}{Caine, J.},
  \bibinfo{author}{Wirth, M.}, \bibinfo{year}{2010}b.
\newblock \bibinfo{title}{Mechanisms of spray formation and combustion from a
  multi-hole injector with e85 and gasoline}.
\newblock \bibinfo{journal}{Combustion and Flame} \bibinfo{volume}{157},
  \bibinfo{pages}{735--756}.
\newblock \DOIprefix\doi{https://doi.org/10.1016/j.combustflame.2009.12.019}.
\bibitem[{Aleiferis and {van Romunde}(2013)}]{Aleiferis2013}
\bibinfo{author}{Aleiferis, P.}, \bibinfo{author}{{van Romunde}, Z.R.},
  \bibinfo{year}{2013}.
\newblock \bibinfo{title}{An analysis of spray development with iso-octane,
  n-pentane, gasoline, ethanol and n-butanol from a multi-hole injector under
  hot fuel conditions}.
\newblock \bibinfo{journal}{Fuel} \bibinfo{volume}{105},
  \bibinfo{pages}{143--168}.
\newblock \DOIprefix\doi{http://dx.doi.org/10.1016/j.fuel.2012.07.044}.
\bibitem[{Badawy et~al.(2022)Badawy, Xu and Li}]{Badawy2022}
\bibinfo{author}{Badawy, T.}, \bibinfo{author}{Xu, H.}, \bibinfo{author}{Li,
  Y.}, \bibinfo{year}{2022}.
\newblock \bibinfo{title}{Macroscopic spray characteristics of iso-octane,
  ethanol, gasoline and methanol from a multi-hole injector under flash boiling
  conditions}.
\newblock \bibinfo{journal}{Fuel} \bibinfo{volume}{307},
  \bibinfo{pages}{121820}.
\newblock \DOIprefix\doi{https://doi.org/10.1016/j.fuel.2021.121820}.
\bibitem[{Brown and York(1962)}]{Brown1962}
\bibinfo{author}{Brown, R.}, \bibinfo{author}{York, J.}, \bibinfo{year}{1962}.
\newblock \bibinfo{title}{Sprays formed by flashing liquid jets}.
\newblock \bibinfo{journal}{AIChE Journal} \bibinfo{volume}{8},
  \bibinfo{pages}{149--53}.
\bibitem[{CMT(2023)}]{CMT2023}
\bibinfo{author}{CMT}, \bibinfo{year}{2023}.
\newblock \bibinfo{title}{Virtual injection rate generator, {CMT-Motores
  T{\'e}r micos, Universitat Polit{\`e}cnica de Val{\`e}ncia}}.
\newblock \URLprefix \url{https://www.cmt.upv.es/}. \bibinfo{note}{{Accessed on
  21.03.2023}}.
\bibitem[{Crowe et~al.(2012)Crowe, Schwarzkopf, Sommerfeld and
  Tsuji}]{Crowe2012}
\bibinfo{author}{Crowe, C.T.}, \bibinfo{author}{Schwarzkopf, J.D.},
  \bibinfo{author}{Sommerfeld, M.}, \bibinfo{author}{Tsuji, Y.},
  \bibinfo{year}{2012}.
\newblock \bibinfo{title}{{Mulitphase Flows with Droplets and Particles}}.
\newblock \bibinfo{publisher}{Taylor \& Francis Group, LLC}.
\bibitem[{Desantes et~al.(2009)Desantes, Pastor, Garc{\'{i}}a-Oliver and
  Pastor}]{Desantes2009}
\bibinfo{author}{Desantes, J.}, \bibinfo{author}{Pastor, J.},
  \bibinfo{author}{Garc{\'{i}}a-Oliver, J.}, \bibinfo{author}{Pastor, J.},
  \bibinfo{year}{2009}.
\newblock \bibinfo{title}{{A 1D model for the description of mixing-controlled
  reacting diesel sprays}}.
\newblock \bibinfo{journal}{Combustion and Flame} \bibinfo{volume}{156},
  \bibinfo{pages}{234--249}.
\newblock \DOIprefix\doi{10.1016/j.combustflame.2008.10.008}.
\bibitem[{Desantes et~al.(2006)Desantes, Payri, Salvador and
  Gil}]{Desantes2006}
\bibinfo{author}{Desantes, J.M.}, \bibinfo{author}{Payri, R.},
  \bibinfo{author}{Salvador, F.J.}, \bibinfo{author}{Gil, A.},
  \bibinfo{year}{2006}.
\newblock \bibinfo{title}{{Development and validation of a theoretical model
  for diesel spray penetration}}.
\newblock \bibinfo{journal}{Fuel} \bibinfo{volume}{22},
  \bibinfo{pages}{87--110}.
\newblock \DOIprefix\doi{10.1016/j.fuel.2005.10.023}.
\bibitem[{Deshmukh et~al.(2022a)Deshmukh, Davidovic, Grenga, Lakshmanan, Cai
  and Pitsch}]{Deshmukh2022b}
\bibinfo{author}{Deshmukh, A.Y.}, \bibinfo{author}{Davidovic, M.},
  \bibinfo{author}{Grenga, T.}, \bibinfo{author}{Lakshmanan, R.},
  \bibinfo{author}{Cai, L.}, \bibinfo{author}{Pitsch, H.},
  \bibinfo{year}{2022}a.
\newblock \bibinfo{title}{A reduced-order model for turbulent reactive sprays
  in compression ignition engines}.
\newblock \bibinfo{journal}{Combustion and Flame} \bibinfo{volume}{236},
  \bibinfo{pages}{111751}.
\newblock \DOIprefix\doi{https://doi.org/10.1016/j.combustflame.2021.111751}.
\bibitem[{Deshmukh et~al.(2022b)Deshmukh, Grenga, Davidovic, Schumacher,
  Palmer, Reddemann, Kneer and Pitsch}]{Deshmukh2022}
\bibinfo{author}{Deshmukh, A.Y.}, \bibinfo{author}{Grenga, T.},
  \bibinfo{author}{Davidovic, M.}, \bibinfo{author}{Schumacher, L.},
  \bibinfo{author}{Palmer, J.}, \bibinfo{author}{Reddemann, M.A.},
  \bibinfo{author}{Kneer, R.}, \bibinfo{author}{Pitsch, H.},
  \bibinfo{year}{2022}b.
\newblock \bibinfo{title}{A reduced-order model for multiphase simulation of
  transient inert sprays}.
\newblock \bibinfo{journal}{Internation Journal of Multiphase Flows}
  \bibinfo{volume}{147}, \bibinfo{pages}{103872}.
\newblock
  \DOIprefix\doi{https://doi.org/10.1016/j.ijmultiphaseflow.2021.103872}.
\bibitem[{Dietzel(2020)}]{Dietzel2020}
\bibinfo{author}{Dietzel, D.R.}, \bibinfo{year}{2020}.
\newblock \bibinfo{title}{Modeling and simulation of flash-boiling of cryogenic
  liquids}.
\newblock Ph.D. thesis.
\bibitem[{Duronio et~al.(2022)Duronio, Mascio, Villante, Anatone and
  Vita}]{Duronio2022}
\bibinfo{author}{Duronio, F.}, \bibinfo{author}{Mascio, A.D.},
  \bibinfo{author}{Villante, C.}, \bibinfo{author}{Anatone, M.},
  \bibinfo{author}{Vita, A.D.}, \bibinfo{year}{2022}.
\newblock \bibinfo{title}{Ecn spray g: Coupled eulerian internal nozzle flow
  and lagrangian spray simulation in flash boiling conditions}.
\newblock \bibinfo{journal}{International Journal of Engine Research}
  \bibinfo{volume}{0}, \bibinfo{pages}{14680874221090732}.
\newblock \DOIprefix\doi{10.1177/14680874221090732}.
\bibitem[{Duronio et~al.(2021)Duronio, Ranieri, Montanaro, Allocca and {De
  Vita}}]{Duronio2021}
\bibinfo{author}{Duronio, F.}, \bibinfo{author}{Ranieri, S.},
  \bibinfo{author}{Montanaro, A.}, \bibinfo{author}{Allocca, L.},
  \bibinfo{author}{{De Vita}, A.}, \bibinfo{year}{2021}.
\newblock \bibinfo{title}{Ecn spray g injector: Numerical modelling of
  flash-boiling breakup and spray collapse}.
\newblock \bibinfo{journal}{International Journal of Multiphase Flow}
  \bibinfo{volume}{145}, \bibinfo{pages}{103817}.
\newblock
  \DOIprefix\doi{https://doi.org/10.1016/j.ijmultiphaseflow.2021.103817}.
\bibitem[{ECN(2020)}]{ECN2020}
\bibinfo{author}{ECN}, \bibinfo{year}{2020}.
\newblock \bibinfo{title}{{Engine Combustion Network (ECN)}}.
\newblock \URLprefix \url{https://ecn.sandia.gov/}. \bibinfo{note}{{Accessed on
  12.10.2020}}.
\bibitem[{Fujimoto et~al.(1997)Fujimoto, Iwami and Senda}]{Fujimoto1997}
\bibinfo{author}{Fujimoto, H.}, \bibinfo{author}{Iwami, Y.},
  \bibinfo{author}{Senda, J.}, \bibinfo{year}{1997}.
\newblock \bibinfo{title}{{Atomization Characteristics of Liquefied n-Butane
  Spray with Flash Boiling Phenomena}}.
\newblock \bibinfo{journal}{International Journal of Fluid Mechanics Research}
  \bibinfo{volume}{24}, \bibinfo{pages}{273--282}.
\newblock \DOIprefix\doi{10.1615/InterJFluidMechRes.v24.i1-3.270}.
\bibitem[{Gemci et~al.(2004)Gemci, Yakut, Chigier and Ho}]{Gemci2004}
\bibinfo{author}{Gemci, T.}, \bibinfo{author}{Yakut, K.},
  \bibinfo{author}{Chigier, N.}, \bibinfo{author}{Ho, T.C.},
  \bibinfo{year}{2004}.
\newblock \bibinfo{title}{Experimental study of flash atomization of binary
  hydrocarbon liquids}.
\newblock \bibinfo{journal}{International Journal of Multiphase Flows}
  \bibinfo{volume}{30}, \bibinfo{pages}{395--417}.
\newblock \DOIprefix\doi{doi:10.1016/j.ijmultiphaseflow.2003.12.003}.
\bibitem[{Germano et~al.(1991)Germano, Piomelli, Moin and Cabot}]{Germano1991}
\bibinfo{author}{Germano, M.}, \bibinfo{author}{Piomelli, U.},
  \bibinfo{author}{Moin, P.}, \bibinfo{author}{Cabot, W.H.},
  \bibinfo{year}{1991}.
\newblock \bibinfo{title}{A dynamic subgrid‐scale eddy viscosity model}.
\newblock \bibinfo{journal}{Physics of Fluids A: Fluid Dynamics}
  \bibinfo{volume}{3}, \bibinfo{pages}{1760--1765}.
\newblock \DOIprefix\doi{10.1063/1.857955}.
\bibitem[{Ghandilou and Taghavifar(2022)}]{Ali2022}
\bibinfo{author}{Ghandilou, A.J.}, \bibinfo{author}{Taghavifar, H.},
  \bibinfo{year}{2022}.
\newblock \bibinfo{title}{New insight into air/spray boundary interaction for
  diesel and biodiesel fuels under different fuel temperatures}.
\newblock \bibinfo{journal}{Biofuels} \bibinfo{volume}{13},
  \bibinfo{pages}{{1087--1101}}.
\newblock \DOIprefix\doi{https://doi.org/10.1080/17597269.2022.2105867}.
\bibitem[{Guo et~al.(2017a)Guo, Ding, Li, Ma, Wang, Xu and Wang}]{Guo2017b}
\bibinfo{author}{Guo, H.}, \bibinfo{author}{Ding, H.}, \bibinfo{author}{Li,
  Y.}, \bibinfo{author}{Ma, X.}, \bibinfo{author}{Wang, Z.},
  \bibinfo{author}{Xu, H.}, \bibinfo{author}{Wang, J.}, \bibinfo{year}{2017}a.
\newblock \bibinfo{title}{Comparison of spray collapses at elevated ambient
  pressure and flash boiling conditions using multi-hole gasoline direct
  injector}.
\newblock \bibinfo{journal}{Fuel} \bibinfo{volume}{199},
  \bibinfo{pages}{125--134}.
\newblock \DOIprefix\doi{https://doi.org/10.1016/j.fuel.2017.02.071}.
\bibitem[{Guo et~al.(2019)Guo, Li, Wang, Zhang and Xu}]{Guo2019b}
\bibinfo{author}{Guo, H.}, \bibinfo{author}{Li, Y.}, \bibinfo{author}{Wang,
  B.}, \bibinfo{author}{Zhang, H.}, \bibinfo{author}{Xu, H.},
  \bibinfo{year}{2019}.
\newblock \bibinfo{title}{Numerical investigation on flashing jet behaviors of
  single-hole gdi injector}.
\newblock \bibinfo{journal}{International Journal of Heat and Mass Transfer}
  \bibinfo{volume}{130}, \bibinfo{pages}{50--59}.
\newblock
  \DOIprefix\doi{https://doi.org/10.1016/j.ijheatmasstransfer.2018.10.088}.
\bibitem[{Guo et~al.(2017b)Guo, Ma, Li, Liang, Wang, Xu and Wang}]{Guo2017a}
\bibinfo{author}{Guo, H.}, \bibinfo{author}{Ma, X.}, \bibinfo{author}{Li, Y.},
  \bibinfo{author}{Liang, S.}, \bibinfo{author}{Wang, Z.}, \bibinfo{author}{Xu,
  H.}, \bibinfo{author}{Wang, J.}, \bibinfo{year}{2017}b.
\newblock \bibinfo{title}{Effect of flash boiling on microscopic and
  macroscopic spray characteristics in optical gdi engine}.
\newblock \bibinfo{journal}{Fuel} \bibinfo{volume}{190},
  \bibinfo{pages}{79--89}.
\newblock \DOIprefix\doi{http://dx.doi.org/10.1016/j.fuel.2016.11.043}.
\bibitem[{Guo et~al.(2018)Guo, Wang, Li, Xu and Wu}]{Guo2018}
\bibinfo{author}{Guo, H.}, \bibinfo{author}{Wang, B.}, \bibinfo{author}{Li,
  Y.}, \bibinfo{author}{Xu, H.}, \bibinfo{author}{Wu, Z.},
  \bibinfo{year}{2018}.
\newblock \bibinfo{title}{Characterizing external flashing jet from single-hole
  gdi injector}.
\newblock \bibinfo{journal}{International Journal of Heat and Mass Transfer}
  \bibinfo{volume}{121}, \bibinfo{pages}{924--932}.
\newblock
  \DOIprefix\doi{https://doi.org/10.1016/j.ijheatmasstransfer.2018.01.042}.
\bibitem[{Hiroyasu and Arai(1980)}]{Hiroyasu1980}
\bibinfo{author}{Hiroyasu, H.}, \bibinfo{author}{Arai, M.},
  \bibinfo{year}{1980}.
\newblock \bibinfo{title}{{Fuel spray penetration and spray angle of diesel
  engines}}.
\newblock \bibinfo{journal}{Trans. JSAE} \bibinfo{volume}{21},
  \bibinfo{pages}{5--11}.
\bibitem[{Hubbard et~al.(1975)Hubbard, Denny and Mills}]{Hubbard1975}
\bibinfo{author}{Hubbard, G.}, \bibinfo{author}{Denny, V.},
  \bibinfo{author}{Mills, A.}, \bibinfo{year}{1975}.
\newblock \bibinfo{title}{{Droplet evaporation: Effects of transients and
  variable properties}}.
\newblock \bibinfo{journal}{International Journal of Heat and Mass Transfer}
  \bibinfo{volume}{18}, \bibinfo{pages}{1003--1008}.
\newblock \DOIprefix\doi{10.1016/0017-9310(75)90217-3}.
\bibitem[{Kale and Banerjee(2019)}]{Kale2019}
\bibinfo{author}{Kale, R.}, \bibinfo{author}{Banerjee, R.},
  \bibinfo{year}{2019}.
\newblock \bibinfo{title}{Understanding spray and atomization characteristics
  of butanol isomers and isooctane under engine like hot injector body
  conditions}.
\newblock \bibinfo{journal}{Fuel} \bibinfo{volume}{237},
  \bibinfo{pages}{191--201}.
\newblock \DOIprefix\doi{https://doi.org/10.1016/j.fuel.2018.09.142}.
\bibitem[{Kawano et~al.(2004)Kawano, Goto, Odaka and Senda}]{Kawano2004}
\bibinfo{author}{Kawano, D.}, \bibinfo{author}{Goto, Y.},
  \bibinfo{author}{Odaka, M.}, \bibinfo{author}{Senda, J.},
  \bibinfo{year}{2004}.
\newblock \bibinfo{title}{Modeling atomization and vaporization processes of
  flash-boiling spray}, in: \bibinfo{booktitle}{SAE 2004 World Congress \&
  Exhibition}, \bibinfo{publisher}{SAE International}.
\newblock \DOIprefix\doi{https://doi.org/10.4271/2004-01-0534}.
\bibitem[{Leach et~al.(2013)Leach, Stone, Fennell, Hayden, Richardson and
  Wicks}]{Leach2013}
\bibinfo{author}{Leach, F.}, \bibinfo{author}{Stone, R.},
  \bibinfo{author}{Fennell, D.}, \bibinfo{author}{Hayden, D.},
  \bibinfo{author}{Richardson, D.}, \bibinfo{author}{Wicks, N.},
  \bibinfo{year}{2013}.
\newblock \bibinfo{title}{Internal Combustion Engines: Performance, Fuel
  Economy and Emissions}. \bibinfo{publisher}{Woodhead Publishing},
  \bibinfo{address}{London}.
\newblock pp. \bibinfo{pages}{193--202}.
\newblock \DOIprefix\doi{https://doi.org/10.1533/9781782421849.5.193}.
\bibitem[{Lefebvre(1988)}]{Lefebvre1988}
\bibinfo{author}{Lefebvre, A.H.}, \bibinfo{year}{1988}.
\newblock \bibinfo{title}{Atomization and Sprays}.
\newblock \bibinfo{publisher}{Hemisphere Publishing Cooperation},
  \bibinfo{address}{New York, USA}.
\bibitem[{Li et~al.(2017)Li, Zhang, Qi and Xu}]{Li2017}
\bibinfo{author}{Li, S.}, \bibinfo{author}{Zhang, Y.}, \bibinfo{author}{Qi,
  W.}, \bibinfo{author}{Xu, B.}, \bibinfo{year}{2017}.
\newblock \bibinfo{title}{Quantitative observation on characteristics and
  breakup of single superheated droplet}.
\newblock \bibinfo{journal}{Experimental Thermal and Fluid Science}
  \bibinfo{volume}{80}, \bibinfo{pages}{{305--312}}.
\newblock \DOIprefix\doi{https://doi.org/10.1016/j.expthermflusci.2016.09.004}.
\bibitem[{Li et~al.(2018a)Li, Guo, Fei, Ma, Zhang, Chen, Feng and
  Wang}]{Li2018b}
\bibinfo{author}{Li, Y.}, \bibinfo{author}{Guo, H.}, \bibinfo{author}{Fei, S.},
  \bibinfo{author}{Ma, X.}, \bibinfo{author}{Zhang, Z.}, \bibinfo{author}{Chen,
  L.}, \bibinfo{author}{Feng, L.}, \bibinfo{author}{Wang, Z.},
  \bibinfo{year}{2018}a.
\newblock \bibinfo{title}{An exploration on collapse mechanism of multi-jet
  flash-boiling sprays}.
\newblock \bibinfo{journal}{Applied Thermal Engineering} \bibinfo{volume}{134},
  \bibinfo{pages}{20--28}.
\newblock \DOIprefix\doi{https://doi.org/10.1016/j.applthermaleng.2018.01.102}.
\bibitem[{Li et~al.(2018b)Li, Guo, Ma, Qi, Wang, Xu and Shuai}]{Li2018a}
\bibinfo{author}{Li, Y.}, \bibinfo{author}{Guo, H.}, \bibinfo{author}{Ma, X.},
  \bibinfo{author}{Qi, Y.}, \bibinfo{author}{Wang, Z.}, \bibinfo{author}{Xu,
  H.}, \bibinfo{author}{Shuai, S.}, \bibinfo{year}{2018}b.
\newblock \bibinfo{title}{Morphology analysis on multi-jet flash-boiling sprays
  under wide ambient pressures}.
\newblock \bibinfo{journal}{Fuel} \bibinfo{volume}{211},
  \bibinfo{pages}{38--47}.
\newblock \DOIprefix\doi{https://doi.org/10.1016/j.fuel.2017.08.082}.
\bibitem[{Li et~al.(2019)Li, Guo, Zhou, Zhang, Ma and Chen}]{Li2019}
\bibinfo{author}{Li, Y.}, \bibinfo{author}{Guo, H.}, \bibinfo{author}{Zhou,
  Z.}, \bibinfo{author}{Zhang, Z.}, \bibinfo{author}{Ma, X.},
  \bibinfo{author}{Chen, L.}, \bibinfo{year}{2019}.
\newblock \bibinfo{title}{Spray morphology transformation of propane, n-hexane
  and iso-octane under flash-boiling conditions}.
\newblock \bibinfo{journal}{Fuel} \bibinfo{volume}{236},
  \bibinfo{pages}{677--685}.
\newblock \DOIprefix\doi{https://doi.org/10.1016/j.fuel.2018.08.160}.
\bibitem[{Miller and Bellan(1999)}]{Miller1999}
\bibinfo{author}{Miller, R.S.}, \bibinfo{author}{Bellan, J.},
  \bibinfo{year}{1999}.
\newblock \bibinfo{title}{{Direct numerical simulation of a confined
  three-dimensional gas mixing layer with one evaporating
  hydrocarbon-droplet-laden stream}}.
\newblock \bibinfo{journal}{Journal of Fluid Mechanics} \bibinfo{volume}{384},
  \bibinfo{pages}{293--338}.
\newblock \DOIprefix\doi{10.1017/S0022112098004042}.
\bibitem[{Mittal et~al.(2014)Mittal, Kang, Doran, Cook and Pitsch}]{Mittal2014}
\bibinfo{author}{Mittal, V.}, \bibinfo{author}{Kang, S.},
  \bibinfo{author}{Doran, E.}, \bibinfo{author}{Cook, D.},
  \bibinfo{author}{Pitsch, H.}, \bibinfo{year}{2014}.
\newblock \bibinfo{title}{{LES of Gas Exchange in IC Engines}}.
\newblock \bibinfo{journal}{Oil {\&} Gas Science and Technology -- Revue d'IFP
  Energies nouvelles} \bibinfo{volume}{69}, \bibinfo{pages}{29--40}.
\newblock \DOIprefix\doi{10.2516/ogst/2013122}.
\bibitem[{Mojtabi et~al.(2014)Mojtabi, Wigley and Helie}]{Mojtabi2014}
\bibinfo{author}{Mojtabi, M.}, \bibinfo{author}{Wigley, G.},
  \bibinfo{author}{Helie, J.}, \bibinfo{year}{2014}.
\newblock \bibinfo{title}{The effect of flash boiling on the atomization
  performance of gasoline direct injection multistream injectors}.
\newblock \bibinfo{journal}{Atomization and Sprays} \bibinfo{volume}{24},
  \bibinfo{pages}{467--493}.
\newblock \DOIprefix\doi{10.1615/AtomizSpr.2014008296}.
\bibitem[{Pastor et~al.(2008)Pastor, {Javier Lopez}, Garcia and
  Pastor}]{Pastor2008}
\bibinfo{author}{Pastor, J.V.}, \bibinfo{author}{{Javier Lopez}, J.},
  \bibinfo{author}{Garcia, J.}, \bibinfo{author}{Pastor, J.M.},
  \bibinfo{year}{2008}.
\newblock \bibinfo{title}{{A 1D model for the description of mixing-controlled
  inert diesel sprays}}.
\newblock \bibinfo{journal}{Fuel} \bibinfo{volume}{87},
  \bibinfo{pages}{2871--2885}.
\newblock \DOIprefix\doi{10.1016/j.fuel.2008.04.017}.
\bibitem[{Patterson and Reitz(1998)}]{Patterson1998}
\bibinfo{author}{Patterson, M.A.}, \bibinfo{author}{Reitz, R.D.},
  \bibinfo{year}{1998}.
\newblock \bibinfo{title}{Modeling the effects of fuel spray characteristics on
  diesel engine combustion and emission}, in: \bibinfo{booktitle}{International
  Congress \& Exposition}, \bibinfo{publisher}{SAE International}.
\newblock \DOIprefix\doi{https://doi.org/10.4271/980131}.
\bibitem[{Pickett et~al.(2011)Pickett, Manin, Genzale, Siebers, Musculus and
  Idicheria}]{Pickett2011}
\bibinfo{author}{Pickett, L.M.}, \bibinfo{author}{Manin, J.},
  \bibinfo{author}{Genzale, C.L.}, \bibinfo{author}{Siebers, D.L.},
  \bibinfo{author}{Musculus, M.P.}, \bibinfo{author}{Idicheria, C.A.},
  \bibinfo{year}{2011}.
\newblock \bibinfo{title}{{Relationship Between Diesel Fuel Spray Vapor
  Penetration/Dispersion and Local Fuel Mixture Fraction}}.
\newblock \bibinfo{journal}{SAE International Journal of Engines}
  \DOIprefix\doi{10.4271/2011-01-0686}.
\bibitem[{Price et~al.(2015)Price, Hamzehloo, Aleiferis and
  Richardson}]{Price2015}
\bibinfo{author}{Price, C.}, \bibinfo{author}{Hamzehloo, A.},
  \bibinfo{author}{Aleiferis, P.}, \bibinfo{author}{Richardson, D.},
  \bibinfo{year}{2015}.
\newblock \bibinfo{title}{Aspects of numerical modelling of flash-boiling fuel
  sprays}, in: \bibinfo{booktitle}{12th International Conference on Engines \&
  Vehicles}, \bibinfo{publisher}{SAE International}.
\newblock \DOIprefix\doi{https://doi.org/10.4271/2015-24-2463}.
\bibitem[{Price et~al.(2016)Price, Hamzehloo, Aleiferis and
  Richardson}]{Price2016}
\bibinfo{author}{Price, C.}, \bibinfo{author}{Hamzehloo, A.},
  \bibinfo{author}{Aleiferis, P.}, \bibinfo{author}{Richardson, D.},
  \bibinfo{year}{2016}.
\newblock \bibinfo{title}{{An Approach to Modeling Flash-Boiling Fuel Sprays
  for Direct-injection Spark-ignition Engines}}.
\newblock \bibinfo{journal}{Atomization and Sprays} \bibinfo{volume}{26},
  \bibinfo{pages}{1--43}.
\newblock \DOIprefix\doi{10.1615/AtomizSpr.2016015807}.
\bibitem[{Price et~al.(2018)Price, Hamzehloo, Aleiferis and
  Richardson}]{Price2018}
\bibinfo{author}{Price, C.}, \bibinfo{author}{Hamzehloo, A.},
  \bibinfo{author}{Aleiferis, P.}, \bibinfo{author}{Richardson, D.},
  \bibinfo{year}{2018}.
\newblock \bibinfo{title}{{Numerical modelling of fuel spray formation and
  collapse from multi-hole injectors under flash-boiling conditions}}.
\newblock \bibinfo{journal}{Fuel} \bibinfo{volume}{221},
  \bibinfo{pages}{{518--541}}.
\newblock \DOIprefix\doi{10.1016/j.fuel.2018.01.088}.
\bibitem[{Price et~al.(2020)Price, Hamzehloo, Aleiferis and
  Richardson}]{Price2020}
\bibinfo{author}{Price, C.}, \bibinfo{author}{Hamzehloo, A.},
  \bibinfo{author}{Aleiferis, P.}, \bibinfo{author}{Richardson, D.},
  \bibinfo{year}{2020}.
\newblock \bibinfo{title}{Numerical modelling of droplet breakup for
  flash-boiling fuel spray predictions}.
\newblock \bibinfo{journal}{International Journal of Multiphase Flow}
  \bibinfo{volume}{125}, \bibinfo{pages}{103183}.
\newblock
  \DOIprefix\doi{https://doi.org/10.1016/j.ijmultiphaseflow.2019.103183}.
\bibitem[{Ranz and Marshall(1952)}]{Ranz1952}
\bibinfo{author}{Ranz, W.E.}, \bibinfo{author}{Marshall, W.R.},
  \bibinfo{year}{1952}.
\newblock \bibinfo{title}{{Evaporation from drops. Parts I {\&} II.}}
\newblock \bibinfo{journal}{Chem. Eng. Progr} \bibinfo{volume}{48},
  \bibinfo{pages}{141--146; 173--180}.
\newblock \DOIprefix\doi{10.1016/S0924-7963(01)00032-X}.
\bibitem[{Ratcliff et~al.(2016)Ratcliff, Burton, Sindler, Christensen, Fouts,
  Chupka and McCormick}]{Ratcliff2016}
\bibinfo{author}{Ratcliff, M.A.}, \bibinfo{author}{Burton, J.},
  \bibinfo{author}{Sindler, P.}, \bibinfo{author}{Christensen, E.},
  \bibinfo{author}{Fouts, L.}, \bibinfo{author}{Chupka, G.M.},
  \bibinfo{author}{McCormick, R.L.}, \bibinfo{year}{2016}.
\newblock \bibinfo{title}{Knock resistance and fine particle emissions for
  several biomass-derived oxygenates in a direct-injection spark-ignition
  engine}.
\newblock \bibinfo{journal}{SAE International Journal of Fuels and Lubricants}
  \bibinfo{volume}{9}, \bibinfo{pages}{59--70}.
\newblock \URLprefix \url{https://doi.org/10.4271/2016-01-0705},
  \DOIprefix\doi{https://doi.org/10.4271/2016-01-0705}.
\bibitem[{Reitz(1987)}]{Reitz1987}
\bibinfo{author}{Reitz, R.D.}, \bibinfo{year}{1987}.
\newblock \bibinfo{title}{{Modeling Atomization Processes in High-Pressure
  Vaporizing Sprays}}.
\newblock \bibinfo{journal}{Atomization and Sprays} \bibinfo{volume}{3},
  \bibinfo{pages}{309--337}.
\bibitem[{Reitz(2013)}]{Reitz2013}
\bibinfo{author}{Reitz, R.D.}, \bibinfo{year}{2013}.
\newblock \bibinfo{title}{Directions in internal combustion engine research}.
\newblock \bibinfo{journal}{Combustion and Flame} \bibinfo{volume}{160},
  \bibinfo{pages}{1--8}.
\newblock \DOIprefix\doi{https://doi.org/10.1016/j.combustflame.2012.11.002}.
\bibitem[{Reitz and Diwakar(1987)}]{Reitz1987a}
\bibinfo{author}{Reitz, R.D.}, \bibinfo{author}{Diwakar, R.},
  \bibinfo{year}{1987}.
\newblock \bibinfo{title}{Structure of high-pressure fuel sprays}, in:
  \bibinfo{booktitle}{SAE International Congress and Exposition},
  \bibinfo{publisher}{SAE International}.
\newblock \DOIprefix\doi{https://doi.org/10.4271/870598}.
\bibitem[{Rusanov(1961)}]{Rusanov1961}
\bibinfo{author}{Rusanov, V.V.}, \bibinfo{year}{1961}.
\newblock \bibinfo{title}{{Calculation of interaction of non--steady shock
  waves with obstacles}}.
\newblock \bibinfo{journal}{J. Comput. Math. Phys. USSR} .
\bibitem[{Saha et~al.(2021)Saha, Deshmukh, Grenga, Bode, Grunewald, Kaya,
  Kirsch, Reddemann, Kneer and Pitsch}]{Saha2021}
\bibinfo{author}{Saha, A.}, \bibinfo{author}{Deshmukh, A.Y.},
  \bibinfo{author}{Grenga, T.}, \bibinfo{author}{Bode, M.},
  \bibinfo{author}{Grunewald, M.}, \bibinfo{author}{Kaya, Y.},
  \bibinfo{author}{Kirsch, V.}, \bibinfo{author}{Reddemann, M.A.},
  \bibinfo{author}{Kneer, R.}, \bibinfo{author}{Pitsch, H.},
  \bibinfo{year}{2021}.
\newblock \bibinfo{title}{{Numerical Modeling of the Flash Boiling
  Characteristics of E-Fuels at Low Ambient Pressure}}, in:
  \bibinfo{booktitle}{International Conference on Liquid Atomization and Spray
  Systems (ICLASS), 30th September - 2nd September},
  \bibinfo{address}{Edinburgh, Scotland UK}.
\bibitem[{Saha et~al.(2023)Saha, Deshmukh, Grenga and Pitsch}]{Saha2023}
\bibinfo{author}{Saha, A.}, \bibinfo{author}{Deshmukh, A.Y.},
  \bibinfo{author}{Grenga, T.}, \bibinfo{author}{Pitsch, H.},
  \bibinfo{year}{2023}.
\newblock \bibinfo{title}{Dimensional analysis of vapor bubble growth
  considering bubble-bubble interactions in flash boiling microdroplets of
  highly volatile liquid electrofuels}.
\newblock \bibinfo{journal}{International Journal of Multiphase Flow}
  \bibinfo{volume}{165}, \bibinfo{pages}{104479}.
\newblock
  \DOIprefix\doi{https://doi.org/10.1016/j.ijmultiphaseflow.2023.104479}.
\bibitem[{Saha et~al.(2022)Saha, Grenga, Deshmukh, Hinrichs, Bode and
  Pitsch}]{Saha2022}
\bibinfo{author}{Saha, A.}, \bibinfo{author}{Grenga, T.},
  \bibinfo{author}{Deshmukh, A.Y.}, \bibinfo{author}{Hinrichs, J.},
  \bibinfo{author}{Bode, M.}, \bibinfo{author}{Pitsch, H.},
  \bibinfo{year}{2022}.
\newblock \bibinfo{title}{Numerical modeling of single droplet flash boiling
  behavior of e-fuels considering internal and external vaporization}.
\newblock \bibinfo{journal}{Fuel} \bibinfo{volume}{308},
  \bibinfo{pages}{121934}.
\bibitem[{Sazhin et~al.(2001)Sazhin, Feng and Heikal}]{Sazhin2001}
\bibinfo{author}{Sazhin, S.}, \bibinfo{author}{Feng, G.},
  \bibinfo{author}{Heikal, M.}, \bibinfo{year}{2001}.
\newblock \bibinfo{title}{{A model for fuel spray penetration}}.
\newblock \bibinfo{journal}{Fuel} \bibinfo{volume}{80},
  \bibinfo{pages}{2171--2180}.
\newblock \DOIprefix\doi{10.1016/S0016-2361(01)00098-9}.
\bibitem[{Schmitz et~al.(2002)Schmitz, Ipp and Leipertz}]{Schmitz2002}
\bibinfo{author}{Schmitz, I.}, \bibinfo{author}{Ipp, W.},
  \bibinfo{author}{Leipertz, A.}, \bibinfo{year}{2002}.
\newblock \bibinfo{title}{{Flash Boiling Effects on the Development of Gasoline
  Direct-Injection Engine Sprays}}.
\newblock \bibinfo{journal}{SAE Journal of Fuels and Lubricants}
  \bibinfo{volume}{111}, \bibinfo{pages}{1025--1032}.
\newblock \URLprefix \url{https://www.jstor.org/stable/44734585}.
\bibitem[{Senda et~al.(1994)Senda, Hojyo and Fujimoto}]{Senda1994}
\bibinfo{author}{Senda, J.}, \bibinfo{author}{Hojyo, Y.},
  \bibinfo{author}{Fujimoto, H.}, \bibinfo{year}{1994}.
\newblock \bibinfo{title}{Modelling of atomization process in flash boiling
  spray}, in: \bibinfo{booktitle}{International Fuels \& Lubricants Meeting \&
  Exposition}, \bibinfo{publisher}{SAE International}.
\newblock \DOIprefix\doi{https://doi.org/10.4271/941925}.
\bibitem[{Senda et~al.(2008)Senda, Wada, Kawano and Fujimoto}]{Senda2008}
\bibinfo{author}{Senda, J.}, \bibinfo{author}{Wada, Y.},
  \bibinfo{author}{Kawano, D.}, \bibinfo{author}{Fujimoto, H.},
  \bibinfo{year}{2008}.
\newblock \bibinfo{title}{Improvement of combustion and emissions indiesel
  engines by means of enhanced mixtureformation based on flash boiling of mixed
  fuel}.
\newblock \bibinfo{journal}{International Journal of Engine Research}
  \bibinfo{volume}{9}, \bibinfo{pages}{15--27}.
\newblock \DOIprefix\doi{10.1243/14680874JER02007}.
\bibitem[{Serras-Pereira et~al.(2010)Serras-Pereira, {van Romunde}, Aleiferis,
  Richardson, Wallace and Cracknell}]{Serras2010}
\bibinfo{author}{Serras-Pereira, J.}, \bibinfo{author}{{van Romunde}, Z.},
  \bibinfo{author}{Aleiferis, P.}, \bibinfo{author}{Richardson, D.},
  \bibinfo{author}{Wallace, S.}, \bibinfo{author}{Cracknell, R.F.},
  \bibinfo{year}{2010}.
\newblock \bibinfo{title}{Cavitation, primary break-up and flash boiling of
  gasoline, iso-octane and n-pentane with a real-size optical direct-injection
  nozzle}.
\newblock \bibinfo{journal}{Fuel} \bibinfo{volume}{89},
  \bibinfo{pages}{2592--2607}.
\newblock \DOIprefix\doi{https://doi.org/10.1016/j.fuel.2010.03.030}.
\bibitem[{She(2010)}]{She2010}
\bibinfo{author}{She, J.}, \bibinfo{year}{2010}.
\newblock \bibinfo{title}{Experimental study on improvement of diesel
  combustion and emissions using flash boiling injection}, in:
  \bibinfo{booktitle}{SAE 2010 World Congress \& Exhibition},
  \bibinfo{publisher}{SAE International}.
\newblock \URLprefix \url{https://doi.org/10.4271/2010-01-0341},
  \DOIprefix\doi{https://doi.org/10.4271/2010-01-0341}.
\bibitem[{Sher and Elata(1977)}]{Sher1977}
\bibinfo{author}{Sher, E.}, \bibinfo{author}{Elata, C.}, \bibinfo{year}{1977}.
\newblock \bibinfo{title}{{Spray Formation from Pressure Cans by Flashing}}.
\newblock \bibinfo{journal}{Ind. Eng. Chem., Process Des. Dev.}
  \bibinfo{volume}{16}, \bibinfo{pages}{237--242}.
\newblock \DOIprefix\doi{https://doi.org/10.1021/i260062a014}.
\bibitem[{Siebers(1999)}]{Siebers1999}
\bibinfo{author}{Siebers, D.L.}, \bibinfo{year}{1999}.
\newblock \bibinfo{title}{{Scaling liquid-phase fuel penetration in diesel
  sprays based on mixing-limited vaporization}}.
\newblock \bibinfo{journal}{SAE Technical Paper 1999-01-0528}
  \DOIprefix\doi{10.4271/1999-01-0528}.
\bibitem[{Sun et~al.(2021a)Sun, Cui, Nour, Li, Hung and Xu}]{Sun2021a}
\bibinfo{author}{Sun, Z.}, \bibinfo{author}{Cui, M.}, \bibinfo{author}{Nour,
  M.}, \bibinfo{author}{Li, X.}, \bibinfo{author}{Hung, D.},
  \bibinfo{author}{Xu, M.}, \bibinfo{year}{2021}a.
\newblock \bibinfo{title}{Study of flash boiling combustion with different fuel
  injection timings in an optical engine using digital image processing
  diagnostics}.
\newblock \bibinfo{journal}{Fuel} \bibinfo{volume}{284},
  \bibinfo{pages}{119078}.
\newblock \DOIprefix\doi{https://doi.org/10.1016/j.fuel.2020.119078}.
\bibitem[{Sun et~al.(2021b)Sun, Cui, Ye, Yang, Li, Hung and Xu}]{Sun2021b}
\bibinfo{author}{Sun, Z.}, \bibinfo{author}{Cui, M.}, \bibinfo{author}{Ye, C.},
  \bibinfo{author}{Yang, S.}, \bibinfo{author}{Li, X.}, \bibinfo{author}{Hung,
  D.}, \bibinfo{author}{Xu, M.}, \bibinfo{year}{2021}b.
\newblock \bibinfo{title}{Split injection flash boiling spray for high
  efficiency and low emissions in a gdi engine under lean combustion
  condition}.
\newblock \bibinfo{journal}{Proceedings of the Combustion Institute}
  \bibinfo{volume}{38}, \bibinfo{pages}{5769--5779}.
\newblock \DOIprefix\doi{https://doi.org/10.1016/j.proci.2020.05.037}.
\bibitem[{Sun et~al.(2020)Sun, Yang, Nour, Li, Hung and Xu}]{Sun2020}
\bibinfo{author}{Sun, Z.}, \bibinfo{author}{Yang, S.}, \bibinfo{author}{Nour,
  M.}, \bibinfo{author}{Li, X.}, \bibinfo{author}{Hung, D.},
  \bibinfo{author}{Xu, M.}, \bibinfo{year}{2020}.
\newblock \bibinfo{title}{Significant impact of flash boiling spray on
  in-cylinder soot formation and oxidation process}.
\newblock \bibinfo{journal}{Energy Fuels} \bibinfo{volume}{34},
  \bibinfo{pages}{10030--10038}.
\newblock \DOIprefix\doi{https://doi.org/10.1021/acs.energyfuels.0c01942}.
\bibitem[{Vanderwege and Hochgreb(1998)}]{Vanderwege1998}
\bibinfo{author}{Vanderwege, B.A.}, \bibinfo{author}{Hochgreb, S.},
  \bibinfo{year}{1998}.
\newblock \bibinfo{title}{The effect of fuel volatility on sprays from
  high-pressure swirl injectors}.
\newblock \bibinfo{journal}{Symposium (International) on Combustion}
  \bibinfo{volume}{27}, \bibinfo{pages}{1865--1871}.
\newblock \DOIprefix\doi{https://doi.org/10.1016/S0082-0784(98)80029-5}.
\bibitem[{{von Kuensberg Sarre} et~al.(1999){von Kuensberg Sarre}, Kong and
  Reitz}]{Sarre1999}
\bibinfo{author}{{von Kuensberg Sarre}, C.}, \bibinfo{author}{Kong, S.c.},
  \bibinfo{author}{Reitz, R.D.}, \bibinfo{year}{1999}.
\newblock \bibinfo{title}{{Modeling the Effects of Injector Nozzle Geometry on
  Diesel Sprays}}.
\newblock \bibinfo{journal}{SAE Technical Paper 1999-01-0912} ,
  \bibinfo{pages}{1--14}\DOIprefix\doi{10.4271/1999-01-0912}.
\bibitem[{Wallis(1969)}]{Wallis1969}
\bibinfo{author}{Wallis, G.B.}, \bibinfo{year}{1969}.
\newblock \bibinfo{title}{{One-Dimensional Two-Phase Flow}}.
\newblock \bibinfo{publisher}{McGraw-Hill, New York}.
\bibitem[{Wan(1997)}]{Wan1997}
\bibinfo{author}{Wan, Y.}, \bibinfo{year}{1997}.
\newblock \bibinfo{title}{{Numerical Study of Transient Fuel Sprays with
  Autoignition and Combustion Under Diesel-Engine Relevant Conditions}}.
\newblock Ph.D. thesis. RWTH Aachen University.
\bibitem[{Wang et~al.(2020)Wang, Qiao, Ju, Sun and Wang}]{Wang2020}
\bibinfo{author}{Wang, J.}, \bibinfo{author}{Qiao, X.}, \bibinfo{author}{Ju,
  D.}, \bibinfo{author}{Sun, C.}, \bibinfo{author}{Wang, T.},
  \bibinfo{year}{2020}.
\newblock \bibinfo{title}{Bubble nucleation, micro-explosion and residue
  formation in superheated jatropha oil droplet: The phenomena of vapor plume
  and vapor cloud}.
\newblock \bibinfo{journal}{Fuel} \bibinfo{volume}{261},
  \bibinfo{pages}{116431}.
\newblock \URLprefix
  \url{https://www.sciencedirect.com/science/article/pii/S0016236119317855},
  \DOIprefix\doi{https://doi.org/10.1016/j.fuel.2019.116431}.
\bibitem[{Wang et~al.(2017a)Wang, Badawy, Wang, Jiang and Xu}]{Wang2017c}
\bibinfo{author}{Wang, Z.}, \bibinfo{author}{Badawy, T.},
  \bibinfo{author}{Wang, B.}, \bibinfo{author}{Jiang, Y.}, \bibinfo{author}{Xu,
  H.}, \bibinfo{year}{2017}a.
\newblock \bibinfo{title}{Experimental characterization of closely coupled
  split isooctane sprays under flash boiling conditions}.
\newblock \bibinfo{journal}{Applied Energy} \bibinfo{volume}{193},
  \bibinfo{pages}{199--209}.
\newblock \DOIprefix\doi{https://doi.org/10.1016/j.apenergy.2017.02.009}.
\bibitem[{Wang et~al.(2017b)Wang, Liu and Reitz}]{Wang2017a}
\bibinfo{author}{Wang, Z.}, \bibinfo{author}{Liu, H.}, \bibinfo{author}{Reitz,
  R.D.}, \bibinfo{year}{2017}b.
\newblock \bibinfo{title}{Knocking combustion in spark-ignition engines}.
\newblock \bibinfo{journal}{Progress in Energy and Combustion Science}
  \bibinfo{volume}{61}, \bibinfo{pages}{78--112}.
\newblock \DOIprefix\doi{https://doi.org/10.1016/j.pecs.2017.03.004}.
\bibitem[{Wang et~al.(2017c)Wang, Ma, Jiang, Li and Xu}]{Wang2017b}
\bibinfo{author}{Wang, Z.}, \bibinfo{author}{Ma, X.}, \bibinfo{author}{Jiang,
  Y.}, \bibinfo{author}{Li, Y.}, \bibinfo{author}{Xu, H.},
  \bibinfo{year}{2017}c.
\newblock \bibinfo{title}{Influence of deposit on spray behaviour under flash
  boiling condition with the application of closely coupled split injection
  strategy}.
\newblock \bibinfo{journal}{Fuel} \bibinfo{volume}{190},
  \bibinfo{pages}{67--78}.
\newblock \DOIprefix\doi{https://doi.org/10.1016/j.fuel.2016.11.012}.
\bibitem[{Wilke(1950)}]{Wilke1950}
\bibinfo{author}{Wilke, C.R.}, \bibinfo{year}{1950}.
\newblock \bibinfo{title}{{A viscosity equation for gas mixtures}}.
\newblock \bibinfo{journal}{The Journal of Chemical Physics}
  \bibinfo{volume}{18}, \bibinfo{pages}{517}.
\newblock \DOIprefix\doi{10.1063/1.1747673}.
\bibitem[{Xi et~al.(2017)Xi, Liu, Jia, Xie and Yin}]{Xi2017}
\bibinfo{author}{Xi, X.}, \bibinfo{author}{Liu, H.}, \bibinfo{author}{Jia, M.},
  \bibinfo{author}{Xie, M.}, \bibinfo{author}{Yin, H.}, \bibinfo{year}{2017}.
\newblock \bibinfo{title}{{A new flash boiling model for single droplet}}.
\newblock \bibinfo{journal}{Fuel} \bibinfo{volume}{107},
  \bibinfo{pages}{{1129--1137}}.
\newblock \DOIprefix\doi{10.1016/j.ijheatmasstransfer.2016.11.027}.
\bibitem[{Xu et~al.(2013)Xu, Zhang, Zeng, Zhang and Zhang}]{Xu2013}
\bibinfo{author}{Xu, M.}, \bibinfo{author}{Zhang, Y.}, \bibinfo{author}{Zeng,
  W.}, \bibinfo{author}{Zhang, G.}, \bibinfo{author}{Zhang, M.},
  \bibinfo{year}{2013}.
\newblock \bibinfo{title}{{Flash Boiling: Easy and Better Way to Generate Ideal
  Sprays than the High Injection Pressure}}.
\newblock \bibinfo{journal}{SAE International Journal of Fuels and Lubricants}
  \bibinfo{volume}{6}, \bibinfo{pages}{137--148}.
\newblock \DOIprefix\doi{10.4271/2013-01-1614}.
\bibitem[{Yamazaki et~al.(1985)Yamazaki, Miyamoto and Murayama}]{Yamazaki1985}
\bibinfo{author}{Yamazaki, N.}, \bibinfo{author}{Miyamoto, N.},
  \bibinfo{author}{Murayama, T.}, \bibinfo{year}{1985}.
\newblock \bibinfo{title}{{The Effects of Flash Boiling Fuel Injection on Spray
  Characteristics, Combustion, and Engine Performance in DI and IDI Diesel
  Engines}}.
\newblock \bibinfo{journal}{SAE Transactions} \bibinfo{volume}{94},
  \bibinfo{pages}{388--395}.
\newblock \URLprefix \url{https://www.jstor.org/stable/44467680}.
\bibitem[{Yang(2017)}]{Yang2017}
\bibinfo{author}{Yang, S.}, \bibinfo{year}{2017}.
\newblock \bibinfo{title}{{Development and validation of a flash boiling model
  for single-component fuel droplets}}.
\newblock \bibinfo{journal}{Atomization and Sprays} \bibinfo{volume}{27},
  \bibinfo{pages}{{963--997}}.
\newblock \DOIprefix\doi{10.1615/AtomizSpr.2017020237}.
\bibitem[{Yang et~al.(2013)Yang, Song, Wang and Yao}]{Yang2013}
\bibinfo{author}{Yang, S.}, \bibinfo{author}{Song, Z.}, \bibinfo{author}{Wang,
  T.}, \bibinfo{author}{Yao, Z.}, \bibinfo{year}{2013}.
\newblock \bibinfo{title}{An experimental study on phenomenon and mechanism of
  flash boiling spray from a multi-hole gasoline direct injector}.
\newblock \bibinfo{journal}{Atomization and Sprays} \bibinfo{volume}{23},
  \bibinfo{pages}{379--399}.
\newblock \DOIprefix\doi{10.1615/AtomizSpr.2013007539}.
\bibitem[{Zeng et~al.(2012)Zeng, Xu, Zhang, Zhang and Cleary}]{Zeng2012}
\bibinfo{author}{Zeng, W.}, \bibinfo{author}{Xu, M.}, \bibinfo{author}{Zhang,
  G.}, \bibinfo{author}{Zhang, Y.}, \bibinfo{author}{Cleary, D.J.},
  \bibinfo{year}{2012}.
\newblock \bibinfo{title}{Atomization and vaporization for flash-boiling
  multi-hole sprays with alcohol fuels}.
\newblock \bibinfo{journal}{Fuel} \bibinfo{volume}{95},
  \bibinfo{pages}{287--297}.
\newblock \DOIprefix\doi{10.1016/j.fuel.2011.08.048}.
\bibitem[{Zeng and Lee(2001)}]{Zeng2001}
\bibinfo{author}{Zeng, Y.}, \bibinfo{author}{Lee, C.F.F.},
  \bibinfo{year}{2001}.
\newblock \bibinfo{title}{{An Atomization Model For Flash Boiling Sprays}}.
\newblock \bibinfo{journal}{Combustion Science and Technology}
  \bibinfo{volume}{169}, \bibinfo{pages}{45--67}.
\newblock \DOIprefix\doi{http://dx.doi.org/10.1080/00102200108907839}.
\bibitem[{Zhao(2010)}]{Zhao2010}
\bibinfo{author}{Zhao, H.}, \bibinfo{year}{2010}.
\newblock \bibinfo{title}{{Advanced Direct Injection Combustion Engine
  Technologies and Development: Gasoline and Gas Engines}}.
\newblock \bibinfo{publisher}{Cambridge: Woodhead}.
\bibitem[{Zhao(2021)}]{Zhihao2010}
\bibinfo{author}{Zhao, Z.}, \bibinfo{year}{2021}.
\newblock \bibinfo{title}{High injection pressure impinging diesel spray
  characteristics and subsequent soot formation in reacting conditions}.
\newblock Ph.D. thesis. Michigan Technical University.
\newblock \DOIprefix\doi{https://doi.org/10.37099/mtu.dc.etdr/1318}.
\bibitem[{Zhou et~al.(2018)Zhou, Lu and Chen}]{Zhou2018}
\bibinfo{author}{Zhou, Z.F.}, \bibinfo{author}{Lu, G.Y.},
  \bibinfo{author}{Chen, B.}, \bibinfo{year}{2018}.
\newblock \bibinfo{title}{Numerical study on the spray and thermal
  characteristics of r404a flashing spray using openfoam}.
\newblock \bibinfo{journal}{International Journal of Heat and Mass Transfer}
  \bibinfo{volume}{117}, \bibinfo{pages}{1312--1321}.
\newblock
  \DOIprefix\doi{https://doi.org/10.1016/j.ijheatmasstransfer.2017.10.095}.
\bibitem[{Zuo et~al.(2000)Zuo, Gomes and Rutland}]{Zuo2000}
\bibinfo{author}{Zuo, B.}, \bibinfo{author}{Gomes, A.M.},
  \bibinfo{author}{Rutland, C.J.}, \bibinfo{year}{2000}.
\newblock \bibinfo{title}{{Modelling superheated fuel sprays and
  vaproization}}.
\newblock \bibinfo{journal}{International Journal of Engine Research}
  \bibinfo{volume}{1}, \bibinfo{pages}{{321--336}}.
\newblock \DOIprefix\doi{10.1243/1468087001545218}.

\end{thebibliography}





\end{document}